\documentclass[reqno]{amsart}
\usepackage{amssymb,stmaryrd}
\usepackage{amsfonts}
\usepackage{graphicx}
\usepackage{amsmath}
\usepackage{physics}

\numberwithin{equation}{section}
\newtheorem{theorem}{Theorem}[section]
\newtheorem{lemma}[theorem]{Lemma}
\newtheorem{proposition}[theorem]{Proposition}
\newtheorem{corollary}[theorem]{Corollary}
\theoremstyle{definition}
\newtheorem{definition}[theorem]{Definition}
\newtheorem{remark}[theorem]{Remark}

\newcommand{\CC}{{\hbox{{$\mathcal C$}}}}
\newcommand{\CD}{{\hbox{{$\mathcal D$}}}}

\newcommand{\CP}{\hbox{{$\mathcal P$}}}

\newcommand{\CO}{\hbox{{$\mathcal O$}}}

\newcommand{\C}{\mathbb{C}}
\newcommand{\R}{\mathbb{R}}
\newcommand{\F}{\mathbb{F}}
\newcommand{\Z}{\mathbb{Z}}
\newcommand{\N}{\mathbb{N}}

\newcommand{\cg}{\hbox{{$\mathfrak g$}}}

\newcommand{\del}{\partial}

\newcommand{\eps}{{\epsilon}}
\newcommand{\tens}{\mathop{{\otimes}}}

\newcommand{\id}{\mathrm{id}}

\newcommand{\<}{\langle}
\renewcommand{\>}{\rangle}

\newcommand{\extd}{{\rm d}}

\def\bicross{{\blacktriangleright\!\!\triangleleft}}

\newcommand{\twoForm}[2]{e^{#1} \otimes e^{#2}}
\newcommand{\sigFun}[2]{ \sigma(\twoForm{#1}{#2}) }

\allowdisplaybreaks[3]

\begin{document}

    \title{Quantum gravity on finite spacetimes and dynamical mass}
    \keywords{noncommutative geometry, quantum gravity, discrete gravity, quantum spacetime, quantum group, FLRW, black hole, Standard model, scalar field}

    \subjclass[2000]{Primary 81R50, 58B32, 83C57}
    \thanks{Ver 1. The first author was partially supported by CONACyT (M\'exico)}

    \author{J. N. Argota-Quiroz and S. Majid}
    \address{Queen Mary University of London\\
        School of Mathematical Sciences, Mile End Rd, London E1 4NS, UK}

    \email{j.n.argotaquiroz@qmul.ac.uk, s.majid@qmul.ac.uk}

    \begin{abstract}We review quantum gravity model building using the new formalism of `quantum Riemannian geometry' to construct this on finite discrete spaces and on fuzzy ones such as matrix algebras. The  formalism starts with a `differential structure' as a bimodule $\Omega^1$ of differential 1-forms over the coordinate algebra $A$, which could be noncommutative. A quantum metric is a noncommutative rank (0,2) tensor in $\Omega^1\tens_A\Omega^1$, for which we then search for a quantum Levi-Civita connection (this is no longer unique or guaranteed). We outline the three models which have so far been constructed in this formalism, commonalities among them, and issues going forward.  One commonality is  a uniform nonzero variance of metric expectation values in the strong gravity limit. We also outline and discuss the construction of quantum FLRW cosmology and black-hole backgrounds using quantum Riemannian geometry and other recent results. Among new results, we perform a Kaluza-Klein type analysis where we tensor classical spacetime coordinates with a finite quantum Riemannian geometry and we give an example where a scalar field on the total space appears as a multiplet of scalar fields on spacetime with a spread of dynamically generated masses.
 \end{abstract}
    \maketitle

    \section{Introduction}

    Quantum gravity is a 100 year old problem and often described as the holy grail for theoretical physics. But what if we already had quantum gravity models in a form that we could compute and explore, but at some cost to how we work and think about Riemannian geometry? Here we will describe such a situation using noncommutative algebra as a new way of thinking about Riemannian geometry\cite{BegMa}. We will do this in a self-contained and practical `toolkit' manner and then describe the three quantum gravity models so far constructed using it\cite{Ma:sq,ArgMa1,LirMa}, focusing on first impressions of quantum gravity that we can glean from them.  Also, while we want to convince you that these models are somewhat canonical, will also not hide currently ad-hoc  choices where we lack a deeper theory. So there will be two themes:

\begin{enumerate}\item In the constructed models, what physical questions would we like to ask? What are common features of quantum gravity already suggested by these models?
\item How do we construct more models and what are the issues currently to watch out for?
\end{enumerate}

 The origins of our approach lie in the `quantum spacetime hypothesis' -- that the coordinates of spacetime are in fact noncommutative due to quantum gravity effects. We say more about this in Section~\ref{sechyp} but the key thing is that it opened the door both to using ideas of `noncommutative geometry' already current in the 1980s  and to developing a new more explicit style of noncommutative geometry motivated by the geometry of quantum groups (but not limited to them) and by quantum spacetime. This second approach was and is less deep than Connes' approach\cite{Con} but more amenable to model building by physicists and it is the one that evolved into \cite{BegMa} as described in Section~\ref{secpre}.

\begin{figure}\includegraphics[scale=.6]{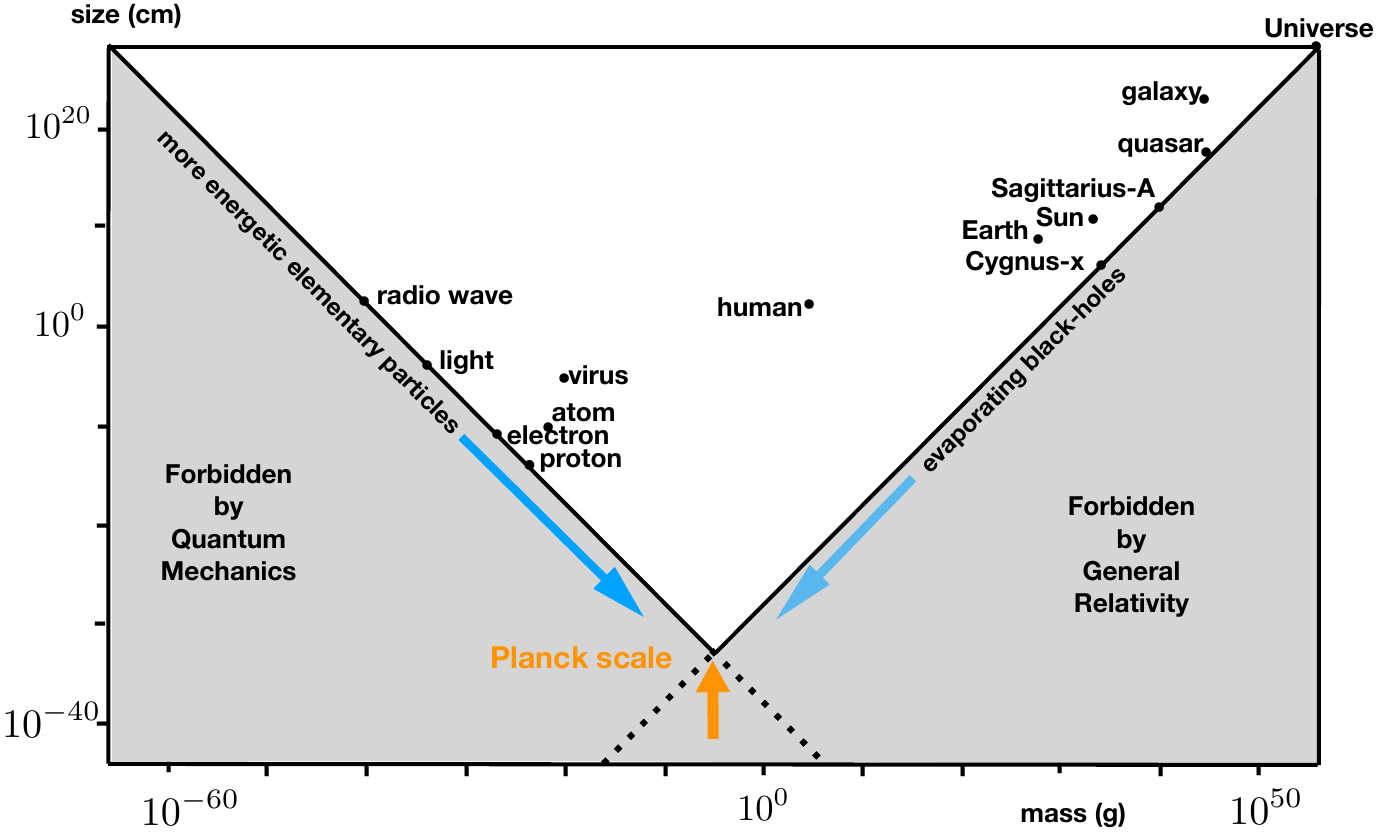}\caption{\label{figplanck} The big picture and the Planck scale from \cite{Ma:pla,BegMa}.}\end{figure}

Now that we have the mathematics of quantum Riemannian geometry (QRG), however, we do no not need to subscribe to the quantum spacetime hypothesis any more. What is more important is that QRG is a more general conception of geometry that includes the continuum at one extreme but also discrete gravity at the other, where the algebra of functions $A$ is finite dimensional. This allows us to address, head-on, the key problem of infinities in quantum field theory, and the extreme ones in the non-renormalisability of quantum gravity.  We recall the fundamental nature of the problem of quantum gravity as the collision of quantum theory and GR  as shown in Figure~\ref{figplanck} \cite{Ma:pla}. The left slope expresses that a particle has Compton wavelength inversely proportional to the mass-energy as part of wave-particle duality in quantum theory. Everything to the left is forbidden by quantum mechanics. The right slope is the line of black holes, where the radius of a black hole is proportional to its mass. Everything to the right is forbidden by GR. In both cases, this is within conventional physical thinking, i.e. {\em} if we take QM and GR at face value then we have ``boxed ourselves' into this triangular region with ourselves somewhat in the middle. The problem then is that probing smaller and smaller distances by elementary particles means we move down the left slope and need particles of smaller wavelength and hence higher mass-energy per quantum and eventually, at $\lambda_P\sim 10^{-33}$cm, the curvature caused by this mass energy is so much as to form a black hole. So distances can never be resolved below this scale and should not be assumed in a theory of physics. Doing so in the case of the quantum modes of gravity itself leads to non-renormalisable infinities. Similarly, coming down the right slope as a black hole evaporates, it eventually reaches the scale where fully quantum effects take over and could well form some kind of hybrid quantum-gravity residue in a theory of quantum gravity. For the same reasons, we do not know the physics right at the centre of a black hole nor in the first $10^{-44}$s from the hypothetical Big Bang.

\begin{figure}\includegraphics[scale=.8]{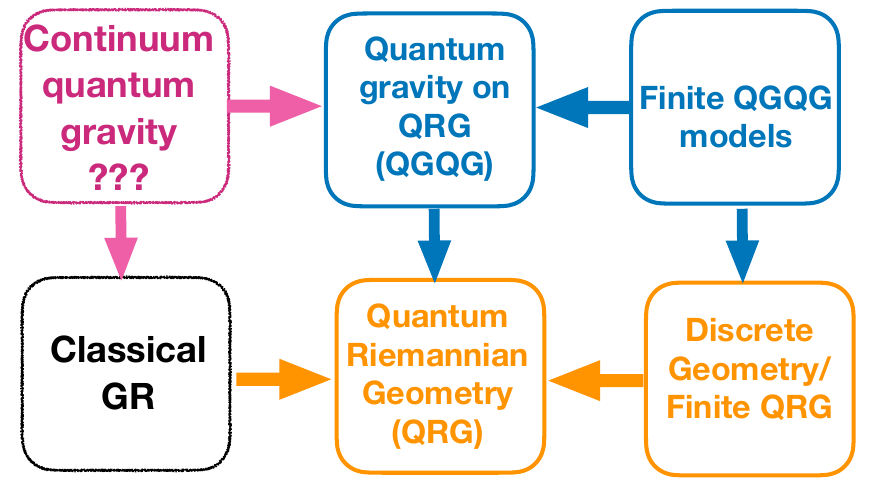}\caption{\label{figqgqg} Embedding continuum quantum gravity (top left)  into a more general QGQG (too middle) and then restricting to finite quantum geometry models as in \cite{Ma:sq,ArgMa1,LirMa,ArgMa3}. The bottom middle and right is more or less known\cite{BegMa}.}\end{figure}

Now, because QRG is simply a more general conception of geometry, there is nothing stopping us doing quantum gravity on a QRG, what we propose to call {\em QGQG models} (standing for quantum gravity on quantum geometry). If we believe the quantum spacetime hypothesis then we should in any case do this as a kind of self-consistency of our point of view. But we can also take the view that if we can do quantum gravity on any manifold, why not just do it more generally on any algebra $A$ within this wider conception of geometry? The algebra $A$ could be noncommutative or it could be commutative (but typically with noncommutative differentials) and it could be finite-dimensional. In the latter case, the problematic `functional integrals' over geometric degrees of freedom would become ordinary integrals. This addresses the problem of quantum gravity head on and changes it to something which is relatively well-defined so that existence is no longer the issue, but at the price that we have to work with an unfamiliar language and, in particular, have to carry over the rest of physics to this setting in order to have a meaningful picture. To the extent that we can do this, we will be able to both construct exact such models {\em and}  physically interpret them to understand quantum gravity effects. Beyond this, the theory specialised to deformation examples would be expected to have essential poles in $\lambda_P$ as a deformation parameter, reflecting the problems of continuum quantum gravity. In short, the idea is we embed the problem of continuum quantum gravity onto the larger problem of quantum gravity on quantum spacetime and then restrict it to finite geometry models where everything is relatively finite and constructible, see Figure~\ref{figqgqg}. But we can also view it as a sophisticated regularisaiton technique where the geometric picture is more fully preserved.

How does this compare with other approaches to quantum gravity? There are of course plenty of ways to make baby models of quantum gravity, typically `mini-superspace' models where we isolate some class of metrics with a finite-dimensional degrees of freedom and just quantise these perhaps within some larger scheme. An important example is loop quantum cosmology\cite{Ash}. Or we can do lattice quantum gravity where we replace spacetime by a finite lattice  as an approximation of the continuum, or more sophisticated variants such as \cite{Loll}. The problem is that if we just write down such a scheme in isolation, we will never know what features mean something and what features are merely artefacts of the approximation. The  key difference in our approach is that our finite quantum gravity models are not approximations to anything. They are quantum gravity on spacetimes which just happen to be finite within this bigger conception of geometry. It means that they are, more or less, part of a single functorial framework that applies across the board with the continuum at one end and finite geometries at the other. This coherence requirement imposes lots of structure on the problem and prevents us from doing random ad-hoc things, and likewise aids the transfer of physical questions from the continuum, if we can phrase them sufficiently geometrically and extend them to any QRG, over to the corresponding questions in the finite geometry models.

This is the idea; the state of the art as we shall see is that we still have to feel our way on certain issues with more or less clear answers in the models worked out but without a fully general theory in some respects relating to the lack of a theory of variational calculus. This means that we do not fully understand stress tensors or the Einstein tensor in QRG but we do have a good sense of the Ricci scalar in at least one constructive approach to the Ricci tensor. Here the Riemann curvature $R_\nabla$ is canonical in QRG but is a 2-form valued operator. In order to take a trace we have to lift the 2-form indices to a (0,2) cotensor by means of map $i:\Omega^2\to \Omega^1\tens_A\Omega^1$, where $\Omega^i$ denotes the space of $i$-forms in our setup. In classical geometry, this map would just be the map that views a 2-form as an antisymmetric (0,2) tensor but in quantum geometry it becomes additional data. We then take a trace to define Ricci. We consider this as a `working definition' and in the models constructed so far there {\em is} an obvious lift $i$, so then there is a natural Ricci scalar curvature $R$. Hence, if we have a map $\int: A\to \C$ on the role of `integration over the spacetime' then we have all the ingredients to write down  a functional integral formulation
\begin{equation}\label{qgfunint} \<\CO\>={\int\extd g\ e^{{\imath\over G} \int R} \CO\over \int\extd g\ e^{{\imath\over G} \int R} },\end{equation}
where $G$ is the gravitational coupling constant and where we omit the $\imath$ in the Euclidean version, and $\CO$ is any function of the metric. Here, $\int$ is the other main loose end and, at the moment, we have to go by what looks reasonable e.g. guided by symmetry or other considerations. Classically, in $n$ dimensions, we would take $\int =\int\extd^n x\sqrt{|g|}(\ ) $, so on a discrete quantum geometry we might expect that this becomes a sum over the points $i$ with a measure $\mu_i$ relating to the metric\cite{Ma:sq,Ma:haw}.

We will take this functional integral formulation as the definition of quantum gravity, but without a theory of noncommutative variational calculus we do not claim to have a corresponding Hamiltonian operator algebra formulation. Known quantum gravity models are in Section~\ref{qgqgmodels}, including a discussion of issues and open problems in Section~\ref{secqgqgdis}.
We will also outline known noncommutative background models, where we do now consider quantum gravity itself, see Section~\ref{secback} and its discussion in Section~\ref{secbackdis}.

Section~\ref{seckk} contains some new results relevant to particle physics. Namely, we  revisit (but now in the context of QRG)  Connes' idea to tensor spacetime coordinates by a finite noncommutative geometry as a way to encode the zoo of elementary particles in the standard model\cite{ConMar}. We fully solve the case of $\Z_n$ for the finite quantum geometry as proof of concept that this approach can produce multiplets of fields on spacetime of different masses, and we analyse but do not fully solve the case of a fuzzy sphere for the extra directions.
The article concludes with some final discussion and outlook in Section~\ref{seccon}.

We will typically work in units with $\hbar=c=1$ but keep $G$. For reference, here are some common acronyms and notations.

\begin{itemize}  \item QRG Quantum Riemannian geometry
\item QG  Quantum gravity
\item GR  General relativity
\item QGQG Quantum gravity on a quantum geometry
\item QLC  Quantum Levi-Civita connection (see Section~\ref{secpre})
\item $(\Omega,\extd)$ the graded exterior algebra of all differential forms with $\Omega^0=A$ the possibly noncommutative `coordinate algebra'  (see Section~\ref{secpre})
\end{itemize}

 \subsection{The quantum spacetime hypothesis}\label{sechyp}

 In all current approaches to quantum gravity, such as strings, loop variables, spin foams, causal sets, a central problem is how exactly does classical GR emerge at scales $>>\lambda_P$. Here, it seems likely that spacetime cannot be a continuum for a consistent theory of quantum gravity, and because quantum effects generally appear as classical observables becoming noncommutative, this suggests, plausibly, that geometry itself, at least at one order better than the classical limit, should be noncommutative:

 \begin{quote}\sl (Quantum spacetime hypothesis): Spacetime coordinates and/or their differentials are noncommutative as a better description of the real world that incorporates some quantum gravity effects.
 \end{quote}

 This has been a matter of speculation since the early days of quantum mechanics and an often cited model is \cite{Sny}, but in fact this did not propose a self-contained algebra of spacetime in a modern sense. The first articulation in a modern context was \cite{Ma:pla} where we argued that  phase space should be both curved and noncommutative  and gave toy models based on quantum groups. Here $x^i$ commmuted, $p_i$ mutually non-commuted corresponding to curvature and $[x^i,p_j]$ non-commute corresponding to quantum mechanics. The bigger picture however, was that the division into position and momentum should be arbitrary and in particular reversible, so one could also have models where $x^i$ noncommute and $p_i$ commute. Such models were discussed explicitly in \cite{MaRue}, where
 \begin{equation}\label{bicross} [x^i,t]=\imath\lambda_P x^i\end{equation}
 and $p_i$ along with some some additional rotation generators formed a quantum Poincar\'e group $\C(\R^3\rtimes \R)\bicross U(so_{1,3})$ isomorphic with $\lambda_P={1\over\kappa}$ to a Hopf algebra proposed by contraction of a q-deformation quantum group in \cite{Luk}. In a modern context, we  prefer deformation  parameters that tend $\lambda_P\to 0$ in the classical limit in the dimensionful case, or $q\to 1$ in the dimensionless case. The Lorentz action on momentum space in the bicrossproduct model is non-linear and deformed in such a way that $|\vec p|<{1\over\lambda_P}$ where $\vec p$ is the spatial momentum, a `doubly special' feature that inspired some authors to consider it as its own starting point (but without convincing other models beyond this already known model).

 Another model that appeared in a similar time was \cite{DFR} similar to the Snyder model but with the Lorentz group un-deformed. It was also argued by t'Hooft \cite{Hoo} that in 2+1 quantum gravity, spacetime could acquire $U(su_2)$ or angular momentum commutation relations
 \begin{equation}\label{spin}  [x^\mu,x^\nu]=2\imath\lambda_P \eps_{\mu\nu\rho}x^\rho\end{equation}
 and this was justified as a `fuzzy $\R^3$' in \cite{BatMa}, which showed that there was a natural quantum group $D(U(su_2))$ acting in this quantum spacetime, and described a suitable covariant differential structure on it. See also many subsequent works including \cite{FreMa}. Indeed, it as explained in  \cite{Ma:dua} that one can also think of noncommutativity of spacetime as curvature in momentum space, which is a classical but curved $SU_2$ in this model (\ref{spin}), and a classical and in some sense curved $\R^3\rtimes \R$ in the bicrossproduct model (\ref{bicross}). Eventually, \cite{MaOse} showed that these two models are, rather surprisingly, related by a Drinfeld twist and hence in some sense equivalent.

 Another modern origin of the quantum spacetime hypothesis was a  proposal  in \cite{Ma:qreg} that this could be Nature's way of  `self-regularisation' where infinities now appear as poles $1/(q-1)$ in q-deformation models. This could be just a mathematical tool where we eventually set $q\to 1$, or it could be that $q\ne 1$ is actually physical. Current thinking is that it could be the latter, given that 2+1 quantum gravity {\em with} cosmological constant $\Lambda$ is based on quantum symmetry $D(U_q(su_2))$, where we use the q-deformed $U(su_2)$ in place of (\ref{spin}) and
 \begin{equation}\label{qlambda} q=e^{-\imath{\lambda_P\over \lambda_C}}\end{equation}
 where $\lambda_P$ is the Planck scale and $\lambda_C=1/\sqrt{|\Lambda|}$ is the length scale associated to the cosmological constant. We refer to  \cite{MaSch} for a fuller discussion.

 Finally, note that if we {\em only} care about $O(\lambda_P)$ corrections to GR then this is a Poisson-level theory worked out in \cite{BegMa:poi} as a valid branch of `first order quantum gravity' which bears the same relation to the full quantum gravity (or rather to those bits of quantum gravity modelled by the quantum spacetime hypothesis) as classical mechanics does to quantum mechanics.

 \subsection{Classification of quantum gravity on quantum geometry (QGQG) models}

 We identify four types of quantum gravity models using such quantum geometries:

 \begin{enumerate}\item Type I (discrete) quantum gravity models.  The algebra $A$ is functions on a directed graph. Here the 1-forms are necessarily noncommutative and spanned as a vector space by the arrows of the graph. Examples are in \cite{Ma:sq,ArgMa1,ArgMa3}
 \item Type II (fuzzy) quantum gravity models. The algebra $A$ has trivial centre and there is a central basis of 1-forms over the algebra. An example is the fuzzy sphere  $\C_\lambda[S^2]$ in \cite{LirMa}.
 \item Type III (deformation) quantum gravity models where there is a continuous deformation parameter. E.g. on  $A=\C_q[S^2]$ the $q$-deformed sphere and the bicrossproduct model (\ref{bicross}) where $q\to 1$ or $\lambda_P\to 0$ is the classical limit of the geometry.
 \item Type IV quantum gravity models. Everything else.
\end{enumerate}

No actual quantum gravity models in Type III or Type IV yet exist. Models of Type III are {\em not} expected to be full deformations in that continuum quantum gravity may not exist due to nonrenormalisable divergences, which means we expect that some quantities  critically involve poles, i.e. $1\over\lambda_P$ or $1/(q-1)$.

\section{Outline of the formalism of QRG} \label{secpre}

It is quite important that our geometric constructions are not ad-hoc but part of a general framework which applies
to most unital algebras and is then restricted to finite-dimensional ones, such as on a finite graph\cite{Ma:gra}. On the other hand, we do not want to burden the reader with the full generality of the theory and so we give only the bare bones at this general level, for orientation.  Details are in \cite{BegMa}. So if you only want the discrete gravity on graphs case then skim this and then go to Section~\ref{secpoly}.

\subsection{Differential structure} We work with a unital possibly noncommutative algebra $A$ viewed as a `coordinate algebra'. We replace the notion of differential structure on a space by specifying a bimodule $\Omega^1$ of differential forms over $A$. A bimodule means we can multiply a `1-form' $\omega\in\Omega^1$ by `functions' $a,b\in A$ either from the left or the right and the two should associate according to
\begin{equation}\label{bimod} (a\omega)b=a(\omega b).\end{equation}
We also need $\extd:A\to \Omega^1$ an `exterior derivative' obeying reasonable axioms, the most important of which is the Leibniz rule
\begin{equation}\label{leib} \extd(ab)=(\extd a)b+ a(\extd b)\end{equation}
for all $a,b\in A$. We usually require $\Omega^1$ to extend to forms of higher degree to give a graded algebra  $\Omega=\oplus\Omega^i$ (where associativity extends the bimodule identity (\ref{bimod}) to higher degree). We also require $\extd$ to extend to $\extd:\Omega^i\to \Omega^{i+1}$ obeying a graded-Leibniz rule with respect to the graded product $\wedge$ and $\extd^2=0$. This much structure is common to most forms of noncommutative geometry, including \cite{Con} albeit there it is not a starting point. In our constructive approach, this `differential structure' is the first choice we have to make in model building once we fixed the algebra $A$. We require that $\Omega$ is generated by $A,\extd A$ as it would be classically.

\begin{definition} We say that $A$ is a {\em differential algebra} if it is an algebra equipped with at least $(\Omega^1,\extd)$ if not a full exterior algebra $(\Omega,\extd)$. We also refer to $(\Omega^1,\extd)$ as a `differential calculus'  on $A$.  \end{definition}

We end this section with  a powerful but purely-quantum concept.

\begin{definition} An exterior algebra $(\Omega,\extd)$ is {\em inner} if there is a 1-form $\theta\in \Omega^1$ such that $\extd =[\theta,\quad\}$ (by which we mean the graded commutator). If we assert this only for $\extd:A\to \Omega^1$, so $\extd a=[\theta, a]$ for all $a\in A$, then we say that $(\Omega^1,\extd)$ is inner.
\end{definition}

This is never possible classically, but is very common in quantum geometry. If anything, it is the typical case and classical geometry is an extreme case.

\subsection{Metrics}

Next, on a differential algebra, we define a metric as an element $g\in \Omega^1\tens_A\Omega^1$ which is invertible in the sense of a map $(\ ,\ ):\Omega^1\tens_A\Omega^1\to A$ which commutes with the product by $A$ from the left or right and inverts $g$ in the sense
\begin{equation}\label{metricinv}((\omega,\ )\tens_A\id)g=\omega=(\id\tens_A(\ ,\omega))g\end{equation}
 for all 1-forms $\omega$. The $\tens_A$ says that we identify elements in the tensor product if they differ by moving a factor $a\in A$ across the tensor prodict. For example, $\omega\tens_A a\eta=\omega a\tens_A \eta$ holds in $\Omega^1\tens_A\Omega^1$. By default, we usually require quantum symmetry in the form $\wedge(g)=0$, where we consider the wedge product on 1-forms as a map $\wedge:\Omega^1\tens_A\Omega^1\to \Omega^2$ and apply this to $g$.  In some contexts, such as graphs in Section~\ref{secpoly} there seems be a more useful alternative notion which we call {\em edge symmetric}. If $g$ does not obey some form of symmetry condition then we call it a `generalised quantum metric'

 \begin{lemma}\cite{BegMa:gra}\cite{BegMa}\label{lem met} If $g$ is a quantum metric or generalised quantum metric then it commutes with functions, $[g,a]=0$.
\end{lemma}
This would not be a restriction in the classical case but in quantum geometry it can be very restrictive if the calculus is very noncommutative. It is an interesting problem how best to relax this, but it also makes sense to live with this in the first instance as it follows from the above very natural assumptions. It  means that quantum gravity under this precise definition could be part of a more general theory with a flabbier notion of quantum metric, but till then it just makes our life easier to have fewer metrics to quantise.

\subsection{Quantum Levi-Civita connection}

Finally, we need the notion of a connection. A left connection on $\Omega^1$ is a linear map $\nabla :\Omega^1\to \Omega^1\tens_A\Omega^1$ obeying a left-Leibniz rule
\begin{equation}\label{connleib}\nabla(a\omega)=\extd a\tens_A \omega+ a\nabla \omega\end{equation}
for all $a\in A, \omega\in \Omega^1$. This might seem mysterious, but if we think of a map $X:\Omega^1\to A$ that commutes with the right action by $A$ as a `vector field' then we can evaluate $\nabla$ to an operator $\nabla_X=(X\tens_A\id)\nabla:\Omega^1\to \Omega^1$ which classically is then a usual covariant derivative on $\Omega^1$. There is a similar notion for a connection on a general `vector bundle' expressed algebraically. Moreover, when we have both left and right actions of $A$ forming a bimodule as we do here, we say that a left connection is a {\em bimodule connection}\cite{DVM,BegMa} if there also exists a bimodule map $\sigma$ such that
\begin{equation}\label{sigma} \sigma:\Omega^1\tens_A\Omega^1\to \Omega^1\tens_A\Omega^1,\quad \nabla(\omega a)=(\nabla\omega)a+\sigma(\omega\tens_A\extd a)\end{equation}
for all $a\in A, \omega\in \Omega^1$.  The map $\sigma$, if it exists, is unique, so this is not additional data but a property that some connections have.  The key thing is that bimodule connections extend automatically to tensor products  as
\begin{equation} \nabla(\omega\tens_A\eta)=\nabla\omega\tens_A\eta+(\sigma(\omega\tens_A(\ ))\tens_A\id)\nabla\eta\end{equation} for all $\omega,\eta\in \Omega^1$, so that metric compatibility now makes sense as $\nabla g=0$. A connection is called  QLC or `quantum Levi-Civita' if it is  metric compatible and the torsion also vanishes, which in our language amounts to $\wedge\nabla=\extd$ as equality of maps $\Omega^1\to \Omega^2$.

 \begin{theorem}\cite{Ma:gra}\cite[Prop. 8.11]{BegMa}\label{innercon}  Let $\Omega$ be inner.
 \begin{enumerate}
 \item A connection on $\Omega^1$ has the form
 \[ \nabla=\theta\tens(\ )-\sigma((\ ) \tens\theta)+\alpha\]
 for two bimodule maps  $\alpha:\Omega^1\to \Omega^1\tens_A\Omega^1$, $\sigma:\Omega^1\tens_A\Omega^1\to \Omega^1\tens_A\Omega^1$.
\item $\nabla$  torsion free is equivalent to $\wedge\alpha=0$ and $\wedge\sigma=-\wedge$.
\item $\nabla g=0$ is equivalent to
\begin{equation*} \theta\tens g + (\id\tens\alpha)g + \sigma_{12}(\id\tens(\alpha - \sigma(\tens\theta)))g = 0. \end{equation*}
\end{enumerate}
\end{theorem}

Hence, in the inner case, the moduli of all connections that we have to search over is equivalent to a pair of bimodule maps $(\alpha,\sigma)$.  Note that the condition (3) in Theorem~\ref{innercon} is {\em quadratic} in $\sigma$, so QLCs will often occur in pairs which in a deformation context could entail one with a classical limit and one `deep quantum' one without a classical limit\cite{BegMa:gra}.

\subsection{Curvature}

\begin{figure}\[\includegraphics[scale=0.8]{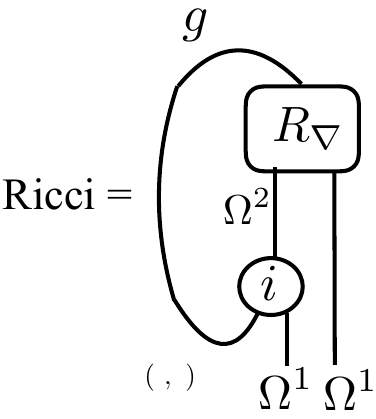}\] \caption{Current `working definition' of Ricci as a trace of Riemann lifted by $i$. Read down the page.\label{figricci}}
\end{figure}

We also have a Riemannian curvature for any connection,
\begin{equation}\label{curv} R_\nabla=(\extd\tens_A\id-\id\wedge\nabla)\nabla:\Omega^1\to \Omega^2\tens_A\Omega^1,\end{equation} where classically one would interior product the first factor against a pair of vector fields to get an operator on 1-forms. Ricci requires more data and the current state of the art (but probably not the only way) is to introduce a lifting bimodule map
\begin{equation}\label{lifting} i:\Omega^2\to\Omega^1\tens_A\Omega^1.\end{equation}
 Applying this to the left output of $R_\nabla$, we are then free to `contract' by using the metric and inverse metric to define ${\rm Ricci}\in \Omega^1\tens_A\Omega^1$ as in Figure~\ref{figricci}. The associated Ricci scalar and the geometric quantum Laplacian are
\begin{equation}\label{LapA} R=(\ ,\ ){\rm Ricci}\in A,\quad \square=(\ ,\ )\nabla\extd: A\to A\end{equation}
defined again along lines that generalise these classical concepts to any algebra with differential structure, metric and connection. Note that in \cite{BegMa}, the scalar curvature is denoted by $S$, but due to the growing importance of entropy in quantum gravity we will now reserve $S$ for that. Also, we are writing $\square$ rather than $\Delta$ as in \cite{BegMa} since, by default, we think of $A$ as spacetime. 

\begin{remark}\label{warn} {\sl (Warning)} The way that Ricci and hence $R$ are defined has a classical limit which is  $-{1\over 2}$ of the usual values. This is because of the way the trace is naturally taken to avoid unnecessary flips and the way the map $i$ is defined as splitting wedge $\wedge\circ i=\id$, which fixes is normalisation. We could put a compensating factor $-2$ into the definition of $R$ but this would be unnatural from the point of view of QRG.
\end{remark}

\subsection{*-structures and integration}

Finally, and critical for physics, are unitarity or `reality' properties. We work over $\C$ but assume that $A$ is a $*$-algebra.  This means $A$ is equipped with an antilinear map $*$ with $*^2=\id$ and $(ab)^*=b^*a^*$ for all $a,b\in A$. In the classical case, real functions on a space would be identified among complex functions as the self-adjoint elements where $a^*=a$.  We require $*$ to extend to $\Omega$ as a graded-anti-involution (reversing order with an extra sign when odd degree differential forms are involved) and to commute with $\extd$.

`Reality' of the metric and of the connection in the sense of being $*$-preserving are imposed as \cite{BegMa:gra,BegMa}
\begin{equation}\label{realgnab} g^\dagger=g,\quad \nabla\circ *= \sigma\circ\dagger\circ \nabla;\quad (\omega \tens_A\eta)^\dagger=\eta^*\tens_A \omega^*,\end{equation}  where $\dagger$ is the natural $*$-operation on $\Omega^1\tens_A\Omega^1$. These `reality' conditions in a self-adjoint basis (if one exists) and in the classical case would ensure that the metric and connection coefficients  are real. For a connection in the inner case with $\theta^*=-\theta$, these conditions become
\begin{equation}\label{inner*con} (\dagger\circ\sigma)^2=\id,\quad \sigma\circ\dagger\circ \alpha= \alpha\circ *.\end{equation}

We will also need a notion of integration $\int: A\to \C$ over the `manifold' underlying $A$.  Classically, it would be given in a local coordinate chart by the Lebesgue measure times a factor $\sqrt{\det(g)}$ but how this is defined for a quantum metric is unclear. Some possible things we might require are as follows. From a quantum mechanical point of view,
\begin{equation}\label{intpos} \overline{\int a}=\int a^*,\quad \int a^*a\ge 0 \end{equation}
with equality if and only if $a=0$. This is a nondegenerate positive linear functional in the sense of $*$-algebras, typically a maximally impure state used to define integration on the algebra. We also want compatibility with the metric and classically this can be done via the divergence of vector fields.  As for Ricci, it is easiest, however, to use the metric to convert vector fields to 1-forms and define the divergence of a 1-form $\omega\in \Omega^1$ via the inverse metric as
\[ {\rm div}(\omega)=(\ ,\ )\nabla\omega.\]
In that case a natural divergence condition on $\int$ motivated by \cite{BegMa:cur} is
\begin{equation}\label{intdiv}  \int a\,  {\rm div}(\omega)= -\int (\extd a,\omega)\end{equation}
for all $a\in A$ and $\omega\in \Omega^1$. We note  that this is compatible with the Leibniz rule:
\begin{align*} \int (ab) {\rm div}(\omega)&=\int a(\ ,\ )(b\nabla\omega)=\int a(\ ,\ )\nabla (b\omega)-\int a(\extd b,\omega)\\
&=-\int (\extd a,b\omega)-\int a(\extd b,\omega)=-\int(\extd(ab),\omega)
\end{align*}
but not necessarily with $*$. For that, it would be natural to impose a further condition
\begin{equation}\label{intsym} \int (\ ,\ )(\id-\sigma)=0.\end{equation}
In fact (\ref{intdiv})-(\ref{intsym}) are too strong in most cases and this is an area for further development, e.g., in connection with quantum geodesics\cite{BegMa:cur}.

\subsection{Algorithm for QGQG model building}\label{secalg}

Given the above, the steps leading to quantum gravity on a quantum spacetime in our approach (a QGQG model) are as follows.

\begin{enumerate} \item Choose a unital $*$-algebra $A$.

\item Make $A$ into a $*$-differential algebra at least to order $\Omega^2$ (we can do without 3-forms or higher for the pure gravity sector.)

\item Choose a class of quantum metrics $g\in \Omega^1\tens_A\Omega^1$ to further quantise, nondegenerate in the sense of having an inverse $(\ ,\ )$ and preferably quantum symmetric or subject to some other similar condition (such as edge-symmetric in the graph case). Describe this moduli explicitly.

\item Solve for the moduli of  QLC's $\nabla: \Omega^1\to \Omega^1\tens_A\Omega^1$ with associated `generalised braiding' $\sigma: \Omega^1\tens_A\Omega^1\to \Omega^1\tens_A\Omega^1$ for each quantum metric in the class is step 3.  Among your solutions, try to identify a canonical choice that works across the whole moduli of metrics.

-- It may be that there is more than one but one is natural (e.g. in having a classical limit).

-- It may be that there is a moduli of QLCs but no preferred one. In that case the quantum gravity theory has to be a  functional integral over the joint moduli of metric-QLC pairs, not just over metrics.

-- Or it there may be that a QLC does not exist for the metrics in your class. In that case go back to step 3.

\item Compute the Riemann curvature for the moduli of QLCs in step 4.

\item Choose a lifting map $i:\Omega^2\to \Omega^1\tens_A\Omega^1$ compatible with $\Omega^2$ and compute {\rm Ricci} with respect to it using the curvatures from step 5. Usually, there will be an obvious choice of $i$.

-- If not, apply some criterion such as that you want {\rm Ricci} to have the same quantum symmetry and $*$-properties as the metric.

-- Or parameterise the possible $i$ as a parameter to your quantum gravity theory.

\item Compute the Ricci scalar $R=(\ ,\ ){\rm Ricci}$ from {\rm Ricci} in step 6.

\item Choose an integration map $\int: A\to \C$ preferably obeying at least the positivity (\ref{intpos}). Similarly to $i$, there will often be an obvious choice or obvious ansatz which you can search among according to what works well.

\item Choose your measure of functional integration on the moduli of quantum metrics as a classical manifold. Again, there will usually be an obvious choice or an obvious ansatz suggested by the classical geometry of the moduli space.

\item {\em Presto!} You have now constructed a candidate for quantum gravity in a functional integral formulation. Explore a bit to see if it looks sensible:

-- Compute some expectation values, cutting off any UV or IR divergences in the metric field strengths with parameters but remembering that only the ratio of integrals enter into the expectation values.

-- If these expectation values still diverge then look at the relative theory of expectation values relative to field expectation values.

-- If the theory does not look very physical then go back and revisit your choices in reverse order (particularly your choice of $\int$ and your choice of $i$).

-- Also look at the relative theory where only fluctuations relative to a mean or background metric are quantised (this tends to have more structure than the fully integrated theory).
\end{enumerate}
Apologies that this is not exactly an algorithm,  but in the examples where it has been tried it seems to work reasonably well. We will see this next.

\section{Existing finite QGQG models}\label{qgqgmodels}

We now show how the toolkit algorithm in Section~\ref{secalg} works for the three examples computed so far in published works\cite{Ma:sq,ArgMa1,LirMa}. Two of them are on graphs where $A$ is the algebra of functions on the nodes and the metric is an assignment of weights to the edges. Here, $A$ is commutative but differentials are intrinsically noncommutative. The other has a very noncommutative $A$ and depends entirely on the above algebraic setup with no classical points in sight.

\subsection{Euclidean discrete (type I) model -- the polygon}  \label{secpoly}

We will model spacetime as a set $X$ of $n$ points with $A$ just the commutative algebra $C(X)$ of functions  on $X$. This is Step 1 of the `algorithm'. Step 2 is to choose a differential structure. Here, it can be shown that $\Omega^1$ are in 1-1 correspondence with directed graphs with vertex set $X$, i.e. with arrows drawn between some vertices (but no duplicated arrows and no self-arrows). We associate a 1-form $\omega_{x\to y}$ to every such arrow $x\to y$ and then $\Omega^1$ as a vector space has all of these as basis.  The bimodule structure that relates the algebra of functions and the differentials, and the exterior derivative are\cite{Ma:gra,BegMa}
\begin{equation}\label{graphbimod}  f.\omega_{x\to y}=f(x)\omega_{x\to y},\quad \omega_{x\to y}.f=f(y)\omega_{x\to y},\quad \extd f=\sum_{x\to y}(f(y)-f(x))\omega_{x\to y}\end{equation}
for any $f\in A$. We see that $f.\omega_{x\to y}\ne \omega_{x\to y}.f$ for generic $f$, so we have a noncommutative differential geometry even though $A$ itself is commutative.  In fact, there is a fundamental reason that noncommutative geometry enters here. This is that a discrete set admits only the zero differential structure in the normal sense and the reason is that a finite difference $f(y)-f(x)$ across an arrow $x\to y$ is an intrinsically {\em bilocal object} -- it does not live at one point of $X$ but exists between two points. In differential geometry, the place where a tensor lives is encoded in the pointwise product of the tensor by any other function, so for a bi-local object we need two products: one pointwise at $x$ and the other pointwise at $y$. These fit naturally with $\Omega^1$ as a bimodule. The bottom line is that the reason that `discrete geometry' on graphs works is that it is really just a special case of QRG.

Step 3 of the `algorithm' is to identify the metrics. These turn out to be of the form \cite{Ma:gra,BegMa}
 \begin{equation}\label{graphmet} g=\sum_{x\to y}g_{x\to y}\omega_{x\to y}\tens_A\omega_{y\to x}\end{equation} with weights $g_{x\to y}\in \R\setminus\{0\}$ for every arrow.  The calculus over $\C$ is compatible with complex conjugation on functions $f^*(x)=\overline{f(x)}$ and $\omega_{x\to y}^*=-\omega_{y\to x}$, from which we see that `reality' of the metric in (\ref{realgnab}) indeed amounts to real metric weights. It is not required but reasonable from the point of view of the physical interpretation, to focus attention on  the {\em edge symmetric} case where $g_{x\to y}=g_{y\to x}$ is independent of the direction.

Step 4 is the hard part -- to solve for a QLC. This actually needs us to go back to Step 2 and say what are the 2-forms. On a general graph, there are some canonical choices starting with a maximal $\Omega_{max}$ \cite{BegMa} and quotients of it. See \cite{Ma:boo,ArgMa3} where this is described and a natural quotient $\Omega_{min}$ is applied to the $A_n$ graph $\bullet$-$\bullet$-$\cdots$-$\bullet$ of $n$ nodes in a line (this is discussed further, below). Once we have fixed the calculus more fully, we note that $\Omega^1$ is inner with $\theta=\sum_{x\to y} \omega_{x\to y}$ so we can use Theorem~\ref{innercon}. Here, $\Omega_{min}$ is also inner.

The situation is somewhat simpler of we identify $X$ with the elements of a finite group $G$ with arrows of the form $x\to xa$ where we right multiply by a generator $a\in \CC$ in a generating set $\CC$ for the group. A graph of this form is called a {\em Cayley graph} and we work with this case for our two discrete examples. The role of $G$ is to define the `manifold' we work over in the same way as one could do GR on $\R^n$ or an a Lie group, i.e. a particularly regular discrete differential structure but with the metric still arbitrary weights on the edges.  In this Cayley graph case, parallel to the Lie group case, there is a basis over the algebra $A$ of left-invariant 1-forms
\[ e^a=\sum_{x\to xa} \omega_{x\to xa};\quad e^a f = R_a(f) e^a,\quad \extd f=\sum_a (R_a(f)- f) e^a;\quad R_a(f)(x)=f(xa)\]
where we sum all the arrows with step given by right multiplication by a fixed $a$. We also state how the bimodule   $\Omega^1$ looks in terms of the right translation operator $R_a$. Being a basis over $A$ means that every 1-form can be written
as $\sum_a \omega_a e^a$  for some coefficients $\omega_a\in A$. Applying this to $\extd f$ defines the partial derivatives
\[( \del_a f)(x)= (R_a f-f)(x)= f(xa)-f(x)\]
in each direction $a\in \CC$. This repackaging of the general graph calculus in the Cayley graph case starts to look much more like familiar formulae on $\R^n$ or on a Lie group. Best of all, there is a canonical choice of $\Omega$ which in the case of $G$ Abelian just says that
\[ \{e_a,e_b\}=0,\quad \extd e_a=0\]
i.e. the $e_a$ form a Grassmann algebra. For non-Abelian $G$ but with $\CC$ a sum of conjugacy classes, one again has a canonical $\Omega$ using an associated braiding to skew-symmetrise. This is a special case of a construction that works for any bicovariant calculus on any Hopf algebra\cite{Wor}. The finite group case notably appeared in the physics literature in \cite{Bre}, with $\Omega^2$ studied in \cite{Ma:non}. We need $\CC$ to be closed under group inversion in order that the calculus admits a quantum metric and a $*$-structure $(e^a)^*=-e^{a^{-1}}$. There is also a canonical map $i:\Omega^2\to \Omega^1\tens_A\Omega^1$ needed for Ricci\cite{BegMa}.

\begin{figure}
    \[\includegraphics[scale=0.7]{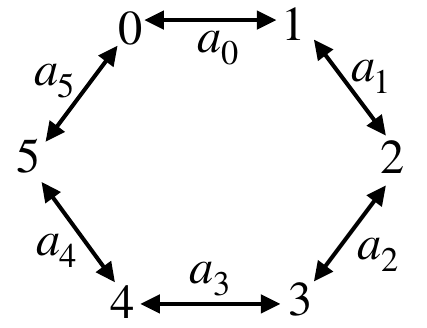}\quad \includegraphics[scale=0.7]{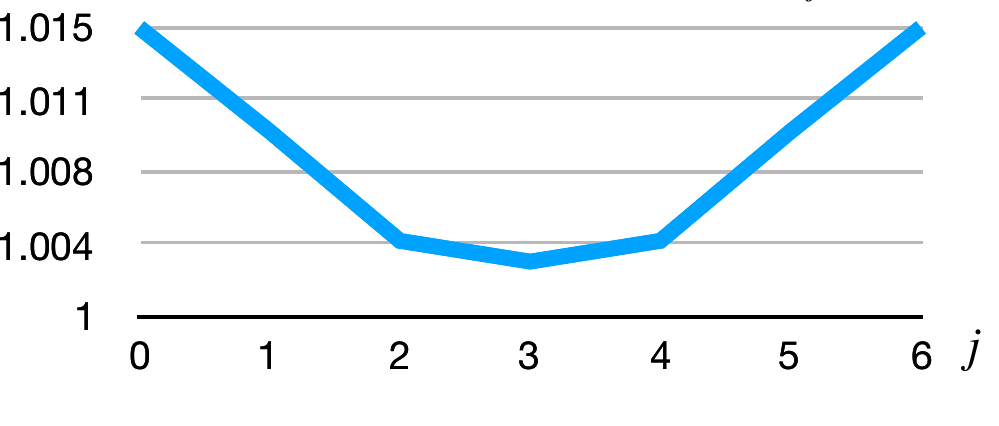}\]
\caption{Graph for the $Z_6$ model and resulting relative correlation functions $\<b_0b_j\>$ from \cite{ArgMa1}. \label{figpoly}}
\end{figure}

To be concrete, we specialise now to $X=G=\Z_n$, a polygon as shown for $n=6$ on the left in Figure~\ref{figpoly}. The vertices are numbered $0,\cdots,n-1$ and we take the generating set to be $\CC={\pm 1}$, i.e. at every node we can step up or step down modulo $n$. The corresponding left-invariant basis is $\{e^+,e^-\}$ with
\[e^+ = \sum_{i=0}^{n-1} \omega_{i \rightarrow i+1}; \quad e^- = \sum_{i=0}^{n-1} \omega_{i \rightarrow i-1},\quad (e^\pm)^{*} = -e^\mp\]
and inner element $\theta = e^+ + e^-$. Next, a metric means weights $a_i$ and we stick to the edge-symmetric case where these are independent of the arrow direction. In terms of the left-invariant 1-forms, this appears as
\begin{equation}\label{metricZ} g=ae^+\tens e^-+R_-(a)e^-\tens e^+,\quad (e^+,e^-) = {1\over a}, \quad (e^-,e^+) = {1\over R_{-}(a)}\end{equation}
where $R_\pm=R_{\pm1}$ and $a\in C(\Z_n)$ is the $g_{+-}$ metric tensor coefficient that encodes the metric data $a(i)=a_i$. We also show the inverse metric as an inner product $(\ ,\ )$ on 1-forms. Because the $e^\pm$ are Grassmann, the canonical lift is just
\[ i(e^+\wedge e^-)={1\over 2}(e^+\tens e^--e^-\tens e^+).\]
Hence we can proceed to Steps 4-7 without any more choices. For Step 4:

\begin{proposition}\cite{ArgMa1}\label{QLCZn} For the $n$-sided polygon with $n\ge 3$, there is a canonical $*$-preserving QLC for an arbitrary metric (\ref{metricZ}), namely determined in Theorem~\ref{innercon} by
    \begin{align*}
        \sigFun{+}{+} = \rho \twoForm{+}{+}; \quad
        \sigFun{+}{-} = \twoForm{-}{+} \\
        \sigFun{-}{+} = \twoForm{+}{-}; \quad
        \sigFun{-}{-} = R_{-}^2(\rho^{-1}) \twoForm{-}{-}
    \end{align*}
and $\alpha=0$, where $\rho = {R_{+}(a)\over a}.$
\end{proposition}
This is unique except for $n=4$ when there is a further 2-parameter moduli of possible $*$-preserving connections. Setting $n=\infty$ is the unique QLC on $\Z$ found in \cite{Ma:haw} for any metric lengths $a_i$ now on the infinite line. The resulting connection and the curvature, Ricci curvature and Ricci scalar curvature (Steps 4,5,6,7) are:
    \begin{align*}
        \nabla e^+ &= (1-\rho) \twoForm{+}{+}, \quad
        \nabla e^- = (1-R_{-}^2(\rho^{-1})) \twoForm{-}{-}, \\
        R_{\nabla}e^+ &= \partial_{-}(\rho) e^+\wedge \twoForm{-}{+}, \quad
        R_{\nabla}e^- = -\partial_{+}(R_{-}^2(\rho^{-1})) e^+\wedge \twoForm{-}{-}, \\
        {\rm Ricci}&= \frac{1}{2}\left(\partial_{-}(R_{-}(\rho)) \twoForm{-}{+} - (\partial_{-}(\rho^{-1}) \twoForm{+}{-} \right), \\
        R &= \frac{1}{2}\left( \frac{\partial_{-}(R_{-}(\rho))}{R_{-}a}-\frac{\partial_{-}(\rho^{-1})}{a} \right).    \end{align*}
Here $\rho$ is a kind of `differential' of the metric, but in a ratio rather than difference sense, and we see that $\nabla$ depends on this while the curvature depends on its further differential.  For Step 8, we need an integration and we take
\[ \int f= \sum_{i=0}^{n-1} a_i f(i)\]
i.e. we weight the translation-invariant integration on the group (the sum over $i$) by a function involving the metric. Classically that measure would be $\sqrt{\det(g)}$ and $a$ plays that role. It is also pointwise positive so that (\ref{intpos}) holds. One does not quite have (\ref{intdiv}),(\ref{intsym}) but we are guided more by the answer. Indeed, with this choice we have the  Einstein-Hilbert-like action
\begin{equation}\label{SgZn} S[g]:=\sum_{\Z_n} a R={1\over 2}\sum_{\Z_n}(R_-\rho)\del_- (R_-\rho)={1\over 2}\sum_{\Z_n}\rho\del_\pm \rho={1\over 4}\sum_{\Z_n}\rho\square_{\Z_n}\rho, \end{equation}
where
\[ (\square_{\Z_n}f)(i):=(\del_++\del_-)(f)(i)=f(i+1)+f(i-1)-2 f(i)\]
 is the usual lattice double-differential on $\Z_n$. In other words, gravity on $\Z_n$ with this measure looks like a lattice scalar field but for a `positive valued' field $\rho$ itself defined as the ratio-differential of the metric $a$. This remarkable feature was already observed for  $\Z$ in \cite{Ma:haw}. The positive nature of the field, in particular, seems to be what distinguishes quantum gravity from scalar quantum field theory.

We are now on Step 9. We can just integrate the action (\ref{SgZn}) over all metric configurations, i.e.
\begin{equation}\label{ZZn}  Z=\int_0^L\extd a_1\cdots\int_0^L\extd a_{n-1}\  e^{{1\over G} S[g]}, \end{equation}
where, for example for $n=3$ (the general pattern is similar),  the action is
\[ S[g]={1\over 2}\left( \frac{a_0}{a_1}+\frac{a_1}{a_2}+\frac{a_2}{a_0}-\frac{a_0^2}{a_2^2} -\frac{a_2^2}{a_1^2}-\frac{a_1^2}{a_0^2}\right),\]
$G$ is a positive coupling constant and $L$ is a cut-off as the integrals are divergent. Continuing with $n=3$, one has finite limits as $L\to \infty$ of
\[ L^{-m-3} \int_0^L\extd a_0\int_0^L\extd a_1\int_0^L\extd a_2\  e^{{1\over G} S[g]} a_{i_1}\cdots a_{i_m}\]
and hence of
\[ L^{-m}\<a_{i_1}\cdots a_{i_m}\>.\]
By symmetry, $\<a_i\>$ is independent of $i$. It follows from these observations that the relative correlation functions where we divide by the relevant power of $\<a_i\>$ have a limit as $L\to \infty$. We can compute them numerically
as functions of $G$ and find
\begin{equation}\label{vevaZn} {\<a_i a_j\>\over \<a_i\>\<a_j\>}\to \begin{cases} {4\over 3} & i=j\\ 1 & i\ne j\end{cases},\quad {\Delta a_i\over \<a_i\>}=\sqrt{\<a_i^2\>-\<a_i\>^2\over \<a_i\>^2}\to {1\over\sqrt{3}}\end{equation}
as $G\to \infty$ (the `deep quantum gravity' regime).

The paper \cite{ArgMa1} does similar calculations for $b_i=a_i/\bar a$, where we divide through by the geometric mean $\bar a=(\prod_i a_i)^{1\over n}$ and then change variables from $a_0,\cdots,a_{n-1}$ to $b_0,\cdots,b_{n-2},\bar a$ with $b_{n-1}=1/(b_0\cdots b_{n-2})$, remembering to include the Jacobean for this change of variables. We can then just omit the $\bar a$ integral as the action, being scale invariant, does not depend on $\bar a$. This time, all our integrals converge and we do not have to take ratios of expectation values (the relative fluctuation fields $b_i$ already have a ratio built into their definition), but the behaviour is qualitatively the same. The correlation functions $\<b_0b_i\>$  for $n=6$ are shown in Figure~\ref{figpoly}. Even though we singled out $b_{n-1}$ in our change of variables, the results only depend on the index of $i$ (the location of the field) modulo $n$.

\begin{figure}
 \[   \includegraphics[scale=.9]{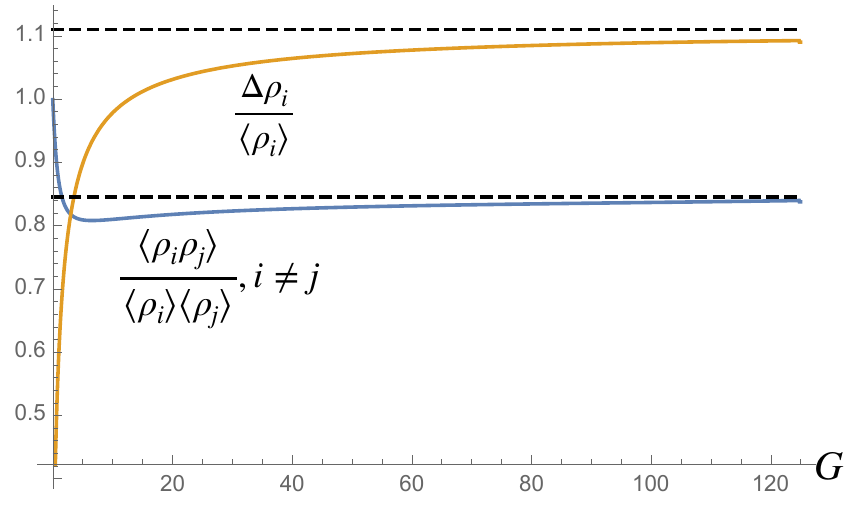}\]
\caption{Treating the $\rho_i$ as the physical variables gives similar features in the $\Z_n$ model, shown here for $n=3$ from \cite{ArgMa3}. \label{figrho}}
\end{figure}

This model can also be used to illustrate a third approach to deal with the scale invariance of the action. This is to regard $\rho_0,\cdots,\rho_{n-1}$ as the effective variables (this does not see an overall scaling of the metric) but note that $\rho_0\cdots\rho_{n-1}=1$ due to the cyclic symmetry. We can regard this as a constrained hyper-surface in $\R^n_{>0}$ and as such this hyper-surface inherits a Riemannian metric $\cg_n$. We can therefore choose $\rho_0,\cdots,\rho_{n-2}$ as the local coordinates and use the Riemannian measure $\int\extd \rho_1\cdots\extd \rho_{n-2}\sqrt{\det(\cg_n)}$ for the functional integral, but the construction is independent of this choice of coordinates. Again, integrals now converge and one has qualitatively the same features of a uniform non-zero relative uncertainty. For example, for $n=3$ \cite{ArgMa1},
\[ {\<\rho_i \rho_j\>\over \<\rho_i\>\<\rho_j\>}\to \begin{cases} 2.23 & i=j\\ 0.845 & i\ne j\end{cases},\quad {\Delta \rho_i\over \<\rho_i\>}\to 1.11\]
as $G\to \infty$, as shown in Figure~\ref{figrho}.

\subsection{Lorentzian discrete (type I) model -- a square}

This was the first QGQG model to be constructed\cite{Ma:sq}, being the first time that the equations
for a QLC could be solved for a significant moduli of quantum metrics -- obviously a prerequisite for quantum gravity.

\begin{figure}
 \[   \includegraphics[scale=1]{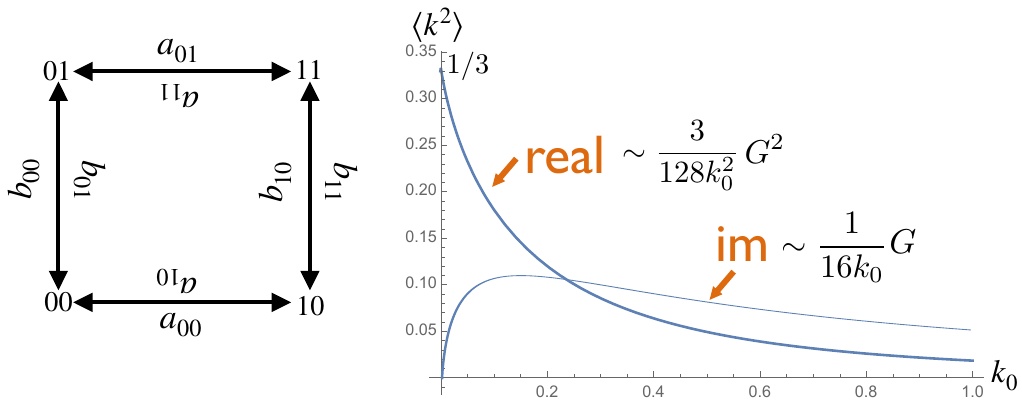}\]
\caption{Finite QGQG Lorentzian model on $\Z_2\times \Z_2$ model \cite{Ma:sq}. On the right is the correlation function in field momentum space for the theory relative to an average value $k_0$ of $a$.  Similarly for $b$. \label{figsquare}}
\end{figure}

The algebra is functions of a set $X$ of 4 points and $\Omega^1$ is chosen as the square. This time we identify the square as the Cayley graph of the group $\Z_2\times\Z_2$, which means a different $\Omega^2$ than the $\Z_4$ case of the  above. This covers Steps 1,2 and we refer to Figure~\ref{figsquare} where we label the vertices by the group in a compact notation $00,01,10,11$. There are 2 basic 1-forms
\[ e^1=\omega_{00\to 10}+\omega_{10\to 00}+\omega_{01\to 11}+\omega_{11\to 01},\quad  e^2=\omega_{00\to 01}+\omega_{01\to00}+\omega_{10\to 11}+\omega_{11\to 10}\]
corresponding to the generators $01$ and $10$ respectively of the group. For Step 3, the general form of a quantum metric is
\begin{equation}\label{squaremet} g=a e^1\tens e^2+be^2\tens e^1,\quad  \del_1 a =\del_2 b=0 \end{equation}
with coefficient functions $a,b\in A$ which correspond to edge values as shown in the figure. The stated condition is for edge symmetry (so that two values at an edge are the same), with $\del_1=R_{10}-\id$ and $\del_2=R_{01}-\id$.  Steps 4 is to solve the QLC equations and we find an $S^1$-moduli of $*$-preserving QLCs labelled by phase parameter. As the $e^i$ are Grassmann, we have a canonical  antisymmetric lift
\[ i(e^1\wedge e^2)={1\over 2}(e^1\tens e^2-e^2\tens e^1)\]
and can find the curvature, Ricci tensor and Ricci scalar for Steps 5-7. Details of these are in \cite{Ma:sq,BegMa}.

For step 8, we take a sum over the four points with measure $|ab|$ in the role of $|\det(g)|$. The merit of this choice of power is that the action then does not depend on the phase parameter in the QLC,
\[ S[g]=\sum_{\Z_2\times\Z_2}|ab|R=(a_{00}-a_{01})^2({1\over a_{00}}+{1\over a_{01}})+ (b_{00}-b_{10})^2({1\over b_{00}}+{1\over b_{10}})  \]
in the Euclidean case where $a,b>0$. This is minimised at the rectangle $a_{00}=a_{01}$ and $b_{00}=b_{10}$. For Step 9, we just take integrals over the four edge weights $a_{00},a_{01},b_{00},b_{10}$. 

In fact, \cite{Ma:sq} proposes to make a `Lorentzian' choice where $a<0$ as the spacelike edges and $b>0$ as the timelike ones. (More precisely, \cite{Ma:sq} puts a minus sign in the metric in the Lorentzian case so that our $a$ is $-a$ there.) It also brings out the physics to make a linear change of variables 
\[ a=-k_0-k_1\psi,\quad b=l_0+l_1\phi,\quad 0< |k_1|<k_0,\quad 0<|l_1|<l_0\]
to momentum space, where
\[ \phi(i,j)=(-1)^i,\quad \psi(i,j)=(-1)^j\]
are plane waves in the two axes directions and are the only modes allowed due to the edge symmetry. The limitations on $k_i$ are to maintain the signature. In these variables,
\[ S[g]={8k_0 k^2\over 1-k^2}- {8l_0 l^2\over 1-l^2},\quad  k:={k_1\over k_0}, \quad l:={l_1\over l_0}.\]
The $k_0,l_0$ enter rather trivially as effective coupling constants and we change $k_1,l_1$ to the relative variables $k,l$. Remembering the Jacobian for this, the partition function is then (up to a constant)
\[ Z= |\Gamma|^2,\quad \Gamma= \int_{-1}^1\extd k \int_0^L \extd k_0\,  k_0e^{{\imath 8 k_0\over G}({k^2\over 1-k^2})},\]
since the vertical theory (with the $l_0,l$) is the complex conjugate of the horizontal theory (with  $k_0,k$). We still need the cutoff to control divergences, but one finds
\[ \<k_0^m\>={2L^m\over m+2},\quad \<k_0^mk^p\>=0\]
for $p>0$. Translating back to $a_{00},a_{01}$, this means
\[ \<a_{00}\>=\<a_{01}\>=-{3\over 2}L,\quad \<a_{00}^2\>=\<a_{01}^2\>=\<a_{00}a_{01}\>={L^2\over 2},\quad {\Delta a_{00}\over |\<a_{00}\>|}={1\over \sqrt{8}}.\]
The theory with the $b$ variables is identical, but without the minus signs, and the theories effectively decouple.

As before, the divergences go away if we look at the relative theory where we indeed treat $k_0,l_0$ as parameters and fail to do their integration. This has again the separated $k,l$ form, but this time
\[ \Gamma= \int_{-1}^1\extd k\, e^{{\imath 8 k_0\over G}({k^2\over 1-k^2})}\]
with $G/k_0$ as the effective gravitational constant $G$. This time, we see rather more structure with
\[ \<a_{00}\>=\<a_{01}\>=-\<k_0(1\pm k)\>=-k_0,\]
\[ \<a_{00}^2\>=\<a_{01}^2\>=k_0^2(1+\<k^2\>),\quad \<a_{00}a_{01}\>=k_0^2(1-\<k^2\>)\]
with $\<k^2\>$ complex and plotted in Figure~\ref{figsquare}. For the weak field limit $G/k_0\to 0$, we get something imaginary which tends to what the same model gives for a scaler field\cite{Ma:sq} but for $G/k_0\to \infty$,  we obtain $\<k^2\>$  tending to a real value of 1/3. Hence
\begin{equation}\label{vecasq} {\<a_{00}^2\>\over\<a_{00}\>^2}\to {4\over 3},\quad  {\<a_{00}a_{01}\>\over\<a_{00}\>\<a_{01}\>}\to {2\over 3},\quad {\Delta a_{00}\over|\<a_{00}\>|}\to {1\over \sqrt{3}}\end{equation}
at large $G/k_0$, which is remarkably similar to our $\Z_3$ results (\ref{vevaZn}) even though the models are very different.

\subsection{Fuzzy  (type II) model -- the fuzzy sphere}\label{secfuzzy} Here, the algebra $A$ has trivial centre and $\Omega^1$ has a central basis $\{s^i\}$ over $A$. We also take $\Omega$ to be the Grassmann algebra on the $\{s^i\}$. This is not the main topic of these notes -- as it is less accessible than the graph case -- so we will only outline the model and refer to  \cite{LirMa} for details.

The unit fuzzy sphere model\cite{LirMa} has $A$ the quantisation of a coadjoint orbit in $su_2^*$ as $U(su_2)$ modulo a constant value of the quadratic Casimir. Specifically, noncommutative `coordinates' $x_i$ obey
\begin{equation}\label{fuzzysph} [x^i,x^j]=2\imath\lambda_P\eps_{ijk}x^k,\quad \sum_i (x^i)^2=1-\lambda_P^2\end{equation}
with physical coordinates obtained by scaling the $x^i$ by a length scale parameter.  When $\lambda_P=1/n$ for $n\in \N$, the $n$-dimensional representation of $U(su_2)$ descends to $A$ and quotient by its kernel gives reduced  fuzzy spheres $c_n[S^2]\cong M_n(\C)$. (In the physics literature, it is these that are often called fuzzy spheres.) The smallest $SU_2$-invariant $\Omega^1$ is 3-dimensional with Grassmann algebra central basis $s^i$ and calculus relations\cite{BegMa},
 \begin{equation}\label{fuzzycalc}  [x^i,s^j]=0,\quad \extd x^i=\eps_{ijk}x^js^k,\quad \extd s^i=-{1\over 2}\eps_{ijk}s^j\wedge s^k.  \end{equation}
Hence the moduli of quantum metrics $g$ becomes identified with the symmetric space $\CP_3$ of $3\times 3$ positive real matrices. This space has a metric $\cg_{P_3}$ with a Riemannian measure which we use in the functional integral in Step 9.  There is a canonical quantum Levi-Civita connection for Step 4 with coefficients that are constant in the algebra just like the metric coefficients. There is a canonical antisymmetric lift
 \[ i(s^i\wedge s^j)={1\over 2}(s^i\tens s^j-s^j\tens s^i)\]
 since the $s^i$ are Grassmann, and one can compute that the scalar curvature\cite{LirMa} is a multiple of 1,
 \begin{equation}\label{Rfuz} R={1\over 2\det(g)}\left({\rm Tr}(g^2)-{1\over 2}{\rm Tr}(g)^2\right)\end{equation}
 for Steps 5-7. The fuzzy sphere has a natural rotationally invariant integral $\int:A\to \C$ but we only need the value on $1$ and we set  $\int 1=|\det(g)|$ for simplicity (one could also consider taking other powers) for Step 8.  The resulting Euclidean quantum gravity theory is\cite{LirMa}
\begin{equation}\label{Zfuzzy} Z=\int_{\CP_3}{\prod_{i\le j}\extd g_{ij}\over |\det g|^2}e^{-{1\over G}({\rm Tr}(g^2)-{1\over 2}{\rm Tr}(g)^2)}\end{equation}
The action is invariant under $SO_3$ rotations of the metric, so if we are only interested in expectation values of rotational invariant operators, it suffices to work with $g={\rm diag}(\lambda_i)$. The integral becomes
\[ Z=\int_\eps^L\prod_i\extd\lambda_i {|(\lambda_1-\lambda_2)(\lambda_2-\lambda_3)(\lambda_3-\lambda_1)|\over\lambda_1^2\lambda_2^2\lambda_3^2}e^{-{1\over 2G}(\lambda_1^2+\lambda_2^2+\lambda_3^2-2(\lambda_1\lambda_2+\lambda_2\lambda_3+\lambda_3\lambda_1)}\]
\begin{equation}\label{vevlam} L^{-m}\<\lambda_{i_1}\cdots\lambda_{i_m}\>\to {3\over 16},\quad  {\<\lambda_i\lambda_j\>\over \<\lambda_i\>\<\lambda_j\>}={16\over 3},\quad {\Delta \lambda_i\over\<\lambda_i\>}=\sqrt{13\over 3}\end{equation}
as $L\to \infty$. We introduced cutoffs to control divergences at both ends, but $\eps\to 0$ can be taken (up to limitations of numerical precision) for the vacuum expectation values, which then appear to be only divergent with $L$. As before, this divergence is controlled by a power law with the result that relative expectation values are well-defined.  Indeed, the model is qualitatively quite similar to the $\Z_n$ and $\Z_2\times\Z_2$ models but with no dependence on the indices in the correlation functions. This is presumably because we are only asking rotationally invariant questions and, in particular, the $i$ index is not a position or direction index but an eigenvalue label. The system can certainly be explored for the expectation values of more geometrically located observables.

\subsection{Discussion of QGQG models}\label{secqgqgdis}

Based on these very different models yet with surprisingly consistent behaviour, we can make some general observations about the emerging flavour of quantum gravity in this approach.

(1) Literally integrating over all metrics is possible as functional integrals are now ordinary integrals which may diverge but are easily controlled. This does, however, typically leads to divergent field expectations. In principle, this should be absorbable in a field renormalisation but at least for now we took a more immediate approach simply to look at what we called relative correlation functions where we divide by powers of the field expectation values to cancel the dependence on the regulator.
It may be overkill, but renormalisation in our models could be explored systematically using modern renormalisation group methods.

(2) The relative correlation functions seem to be associated with nice arithmetics involving rational numbers.

(3) The field expectation values $\<a_i\>$ and $\<\lambda_i\>$ etc are nonzero and in the Euclidean versions are all positive, but constant in the position or other distinguishing index $i$. On any graph, one has a canonical `Euclidean metric' in the notation of \cite{Ma:gra,BegMa} with 1 on every edge and  we obtain a multiple of this for  the expectation values in the Euclidean models.

(4) In all the models, there is a nonzero but uniform (relative) uncertainty in the metric. This is not too surprising but one can speculate  that it  could be indicative of some kind of vacuum energy.

(5) Much of the structure was, however, integrated out and more was revealed if we did a relative quantisation of some kind. We experimented with some different approaches: relative to the geometric mean in the $\Z_n$ model and relative to the usual mean in the vertical/horizontal sectors of the $\Z_2\times \Z_2$ model. We found reality and positivity features similar to the full theory in the $G\to \infty$ `deep quantum gravity' limit.

(6) One could view fixing a constant and looking at fluctuations around it as a kind of `background field method' but this remains to be looked at in other models and developed more systematically.

(7) In one of the models, we allowed negative metric edge weights $a$ and one can call such an edge `spacelike'\cite{Ma:sq} as opposed to `timelike' if positive, for a Lorentzian point of view. Not much inference can be drawn from just one model, but clearly the idea to develop a notion of causal structure at least in the graph case (designating a preferred arrow direction for the timelike edges) and its consequences for the QGQG models should be explored further and potentially related to causal sets as in \cite{Dow}.

(8) One of the less clear elements of the model building was the choice of measure of integration over a QRG needed for the action. In the absence of a general theory, we chose in the discrete theory a sum over the points with weighting something like a power of $\det(g)$, but not exactly $\sqrt{|\det(g)|}$ as one would have classically. This was guided by simplicity of the result, but other choices could be looked at.

(9) We did not cover matter fields, but there is no obstruction to do so. On every QRG one has a canonical scalar Laplacian $\square: A\to A$ as in (\ref{LapA}) which we can use to define an action and then do functional integration. This we considered in \cite{Ma:sq,Ma:haw} in the discrete models there.

Aside from these specific observations, one would generally like to see some of the expected physics of quantum gravity such as entropy/gravity relations, gravitationally induced state reduction, an understanding of the cosmological constant, quantum smoothing of singularities, etc. For this one would need:

\begin{itemize}
\item To transfer of more physical concepts into the QGQG setting.
\item  More sophisticated models (e.g. rather more points) so that one has enough structure to interrogate.
\item More powerful computers or techniques to do the integrals on larger models.
\item To expect the unexpected (be prepared to generalise our framework as needed).
\end{itemize}

On the last point, it has recently been found\cite{ArgMa3} for the $n$-node graph $\bullet$-$\bullet$-$\cdots$-$\bullet$ that we cannot assume edge symmetry. Non-edge symmetric metrics have always been an option, e.g. in \cite{Ma:haw} and studied for $\Z_n$ in \cite{Sit}, but here we are forced to them. Indeed, for a QLC to exist on the $n$-node finite interval, \cite{ArgMa3} shows that we are forced to have a different length travelling in one direction (into the bulk) compared to the other (towards an endpoint). These are still arbitrary (so we can do quantum gravity) but in a fixed ratio. Moreover, these ratios involve $q$-integers $(q^m-q^{-m})/(q-q^{-1})$ where $q^2$ is an $n+1$-th root of unity, even though there is no quantum group currently in sight\cite{ArgMa3}. One would need to see how such features play out in other graphs but this is an example of something unexpected from the  point of view of classical geometry that could indicate entirely novel physics. Quantum gravity on 2 points was previously considered using Connes' spectral triples approach to noncommutative geometry in \cite{Hal}.

\section{FLRW and black hole fixed background models} \label{secback}

In the remainder of these notes, we do not do quantum gravity itself but turn towards noncommutative  curved backgrounds of interest as hypothetical quantum gravity corrections to classical GR according to the quantum spacetime hypothesis in Section~\ref{sechyp}. We can see how physics is modified e.g. through the behaviour of a scaler field. Here, the QRG on a possibly noncommutative coordinate algebra $A$ provides the canonical Laplacian $\square$
in (\ref{LapA}) using a connection $\nabla$ and inverse metric inner product $(\ ,\ )$.

The standard examples such as bicrossproduct or $\kappa$-Minkowski spacetime (\ref{bicross}) and the fuzzy $\R^3$  (\ref{spin}) have their QRG described in \cite{BegMa}. In both cases, the flat QRG needs $\Omega^1$ of one dimension higher in order have expected (quantum group) symmetries, a phenomenon known as the {\em quantum anomaly for differential calculus}. Usually in physics, an anomaly reflects that a symmetry does not get quantised, but this obstruction can often be absorbed by moving to a higher dimensional spacetime. In our case, we can keep the symmetry (perhaps as a quantum one) but the  differential calculus may not deformation-quantise compatibly with it and we can often absorb this obstruction by a higher dimensional $\Omega^1$. In general, this is one of three approaches:
\begin{itemize}
\item Extend $\Omega^1$ with at least one extra dimension, which is often the inner element $\theta$ or some variant $\theta'$.
\item Drop the classical or quantum symmetry or generalise it e.g. with a cocycle.
\item Allow $A$ or at least differential forms $\Omega$ to become nonassociative.
\end{itemize}
Thus,  for a 4D QRG on the bicrossproduct model (\ref{bicross}), one can drop the quantum Poincar\'e symmetry and then there is a natural 4D calculus which, however, forces the metric to have a specific curved form\cite{BegMa:gra,BegMa}. For the third option, one can indeed analyse the obstruction at a Poisson level \cite{BegMa:poi} and find that it corresponds to the non-existence of a flat (contravariant or `pre') connection with the desired symmetry. The natural approach here is to allow this to be curved,  in which case the 1-forms (but not necessarily the coordinates themselves) become nonassociative. A black hole at this level is in \cite{BegMa:poi}.

Here we stick to the cleanest first option and live with the dimension jump. One might think of $\Z_n$ as like a circle, but we saw in Section~\ref{secpoly} that its natural calculus is 2D with basic 1-forms $e^\pm$ and associated partials $\del_\pm:A\to A$. Similarly for the fuzzy sphere, we saw in Section~\ref{secfuzzy} that its natural calculus is 3D with central basic 1-forms $s^i$ and associated partials $\del_i$. We will now look at both FLRW and black hole models in polar coordinates with the angular part replaced by one of these. There are therefore four models that we describe, with details and derivations in \cite{ArgMa1,ArgMa2}.  In all four models, $t,r$ (as applicable) are classical and commute with everything. Likewise,  $\extd t,\extd r$  are classical and graded-commute with everything, and we let $\del_t,\del_r$ be their  associated classical partial derivatives. So all the noncommutativity is confined to the angular sector. We let
\begin{equation}\label{tenss} \omega\tens_s \eta:=\omega\tens\eta + \eta\tens\omega\end{equation}
be a short hand. The papers \cite{ArgMa1,ArgMa2} look more generally, but for these notes we just take the constant Euclidean metric $-e^+\tens_s e^-$ on $\Z_n$. We did not care about the overall sign of the metric before, but a sign is due when noting that $e^+{}^*=-e^-$ and comparing with the classical picture as explained in \cite{ArgMa1}. Likewise for the fuzzy sphere, we take the round metric $s^i\tens s^i$.

We will need notions of divergence and Einstein tensor. In general, these are not known but for the pressent models this is not a problem and the `naive' definitions suffice, namely
\begin{equation}\label{naiveTE} \nabla\cdot T:= ((\ ,\ )\tens\id)\nabla T,\quad {\rm Eins}:={\rm Ricci}-{g\over 2}R, \end{equation}
by analogy with the classical formulae, where $R$ is the scalar curvature.

\subsection{FLRW background with expanding polygon}

We let $a(t)$ be an arbitrary function of $t$, typically increasing but it does not have to be. This was denoted $R(t)$ in \cite{ArgMa1}, but we now reserve that for the scalar curvature. With the notations explained above, the `expanding polygon' FLRW model has the form of quantum metric and ensuing QRG:
\begin{align*}g&=-\extd t\otimes\extd t- a^2(t) e^+\tens_s e^-,\quad (\extd t,\extd t)=-1,\quad (e^\pm,e^\mp)={1\over a^2(t)},\\
    \nabla \extd t &=  a \dot a  e^+\tens_s e^-,\quad \nabla e^\pm=-  {\dot  a \over  a }e^\pm\tens_s\extd t,\\
    R_\nabla e^\pm&=-{\ddot  a \over  a }\extd t\wedge e^\pm\tens\extd t  \pm \left({\dot  a \over  a }\right)^2   a ^2 e^+\wedge e^-\tens e^\pm,\quad
    R_\nabla \extd t=  \ddot  a   a \extd t\wedge e^+\tens_s e^-,\\
    {\rm Ricci}&= {\ddot  a \over  a } \extd t\tens \extd t+{1\over 2}\left({\dot a ^2\over  a ^2}+{ \ddot a  \over  a }  \right)  a ^2 e^+\tens_s e^- ,\quad  R=-2{\ddot  a \over  a } -\left({\dot  a \over  a }\right)^2,\\
      {\rm Eins} &= {\rm Ricci} - {1\over 2}Rg =  - {1\over 2 }\left( {\dot{a}\over a} \right)^2 \extd t\tens\extd t-{ a \ddot a  \over 2} e^+ \tens_s e^-,\\
     \square &= {2\over a^2} \left( \partial_+ + \partial_- \right) - \partial_t^2 \end{align*}
using the naive definition of the Einstein tensor. One can check that $\nabla\cdot {\rm Eins}=0$. For the stress tensor of dust, we take
\[ T=p g+(f+p)\extd t\tens\extd t=f\extd t\tens\extd t- p a^2 e^+\tens_s e^-\]
for pressure and density functions $p,f$. Then solving the Einstein equations ${\rm Eins}+4\pi GT=0$ (remembering the warning in Remark~\ref{warn}) gives
\[  f={1\over 8 \pi G} \left({\dot a\over a}\right)^2,\quad p=-{1\over 8 \pi G}{\ddot a\over a}\]
and also ensures that $\nabla\cdot T=0$ in the form of the continuity equation $\dot f+2(f+p){\dot a\over a}=0$. Finally, the equation of state $p=\omega f$ for a constant $\omega$ gives
\[ a(t)=a_0(1+\sqrt{8\pi G f_0}(1+\omega t))^{1\over 1+\omega}\]
for initial radius and density $a_0,f_0$. It is shown in \cite{ArgMa1} that this agrees with the classical Friedmann equations and its solution on $\R\times\R^2$. Thus, the expansion is like that of a flat 2D space in place of the $\Z_n$ at each $t$, which fits with the QRG of $\Z_n$ with the constant metric being flat and 2-dimensional.

\begin{figure}
    \includegraphics[scale=.7]{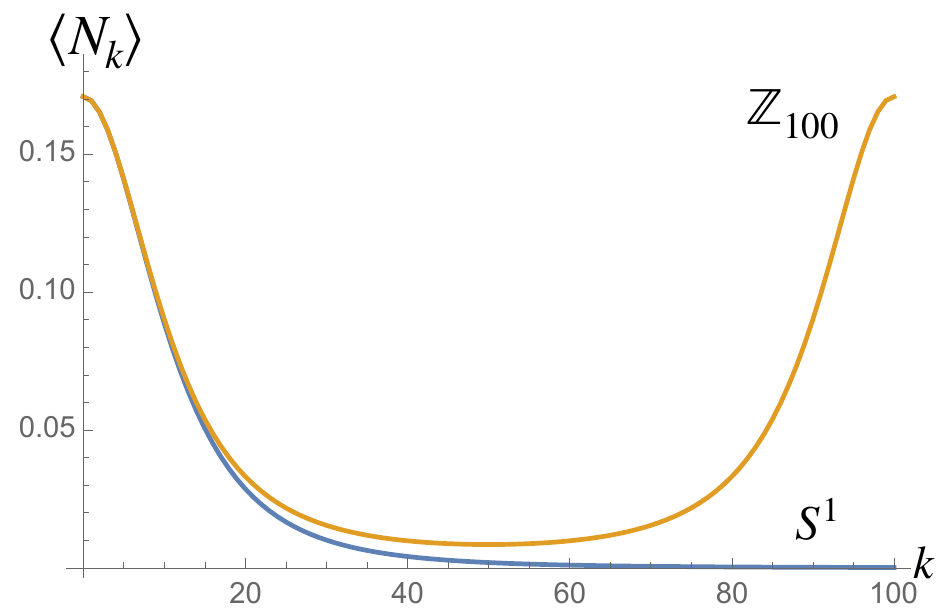}
    \caption{Particle creation in expanding $\Z_{100}$ cosmology against momentum $k$ and compared to $S^1$. Diagram from \cite{ArgMa1} with $a_{in}m=1$, $a_{out}m=\sqrt{5}$ and $\mu=100$.}
    \label{fig:pol-circle}
\end{figure}

We can also do particle creation in this background by solving the Klein-Gordon equation $(-\square+m^2)\phi=0$ in our conventions for $\square$ above\cite{ArgMa1}. Following \cite{Birrel}, we first change to a new time variable $\eta$ such that ${\extd \eta\over\extd t}={1\over a(t)}$ rendering our metric conformally flat, $g= C(\eta)(-\extd \eta\tens\extd \eta - e^+\tens_s e^-)$, where
\[ C(\eta)=a^2(t)= {1\over 2}(a^2_{in}+a^2_{out})+ {\tanh(\mu\eta)\over 2}(a^2_{in}-a^2_{out})\]
is chosen to represent a period of expansion (in the new time variable) from $a_{in}$ to $a_{out}$ and $\mu$ is a parameter for the expansion. Next, we write our Klein-Gordon fields as
\[ \phi(\eta,i)= (2\pi C(\eta))^{-1/2} \sum_{k=0}^{n-1} A_k e^{\frac{2\pi \imath}{n}ik} h_k(\eta) + {h.c.}\]
where  h.c. stands for the hermitian conjugate. Then the Klein-Gordon equation in early and late time reduces  to
\begin{equation*}
    {\extd^2 h_k\over \extd\eta^2}+ w_k^2(\eta)h_k=0, \quad w_k(\eta) = \sqrt{ C(\eta) m^2 + 8\sin^2{\left(\frac{\pi}{n}k\right)  }  }
\end{equation*}
and the values of the `mass on-shell' function for early and late times are
\begin{equation*}
    \label{eq:w_pol}
    w_k^{\rm in}= \sqrt{a^2_{in} m^2+8\sin^2{\left(\frac{\pi}{n}k\right) }} ; \quad w_k^{\rm out}= \sqrt{a^2_{out} m^2+8\sin^2{\left(\frac{\pi}{n}k\right) }}.
\end{equation*}
The rest proceeds in a standard fashion. Writing $w_k^{\pm} = \frac{1}{2}(w_k^{out}\pm w_k^{in})$ and by standard arguments for the quantum field theory with coefficients $A_k,A_k^*$ promoted to operators, the number operator $N_k$ in the initial vacuum state has at late time\cite{Birrel}
\begin{align*}
    \label{eq:N_k}
    \<N_k\> = \frac{\sinh^2{\left(\pi \frac{w^-_k}{\mu}\right)} }{\sinh{(\pi \frac{w_k^{in}}{\mu}  )} \sinh{(\pi \frac{w_k^{out}}{\mu})}}.
\end{align*}
The result in our case is plotted for $n=100$ in Figure~\ref{fig:pol-circle}, where it is compared with the same calculation for $\R\times S^1$ with spatial momentum in terms of $k\in \Z$.  

\subsection{FLRW background with expanding fuzzy sphere}

This works in a similar way with a unit fuzzy sphere for the angular sector, again with expansion factor $a(t)$. The quantum metric and resulting QRG are\cite{ArgMa2}
\begin{align*} g &= - \extd t \tens \extd t + a^2(t) s^i \tens s^i,\quad (\extd t,\extd t)=-1 ,\quad  (s^i,s^j) = {\delta^{ij} \over a^2},\\
    \nabla \extd t &= - a\dot a s^i \tens s^i; \quad \nabla s^i = -\frac{1}{2} \epsilon^{i}{}_{jk}s^j \tens s^k - \frac{\dot{a}}{a} s^i \tens_s \extd t, \\
    R_\nabla \extd t &= -a\ddot a \extd t \wedge s^i\tens s^i, \\
    R_\nabla s^i &= \left(  {1\over 4} \epsilon^{pi}{}_{n}\epsilon_{pkm} - \dot a^2 \delta^{i}{}_{m}\delta_{nk} \right) s^m \wedge s^n \tens s^k + {\ddot a \over a}\extd t \wedge s^i \tens \extd t,\\
    {\rm Ricci} &= -(  \dot a^2 + {1\over 2}a\ddot a +\frac{1}{4})s^i \tens s^i + {3\over 2}{\ddot a \over a} \extd t \tens \extd t,\quad R =-3\left(  \frac{\dot{a}^2}{a^2} + \frac{\ddot{a}}{a} +\frac{1}{4a^2} \right),\\
   {\rm Eins} &= \left(  \ddot{a} + \frac{1}{2}\dot{a}^2 +\frac{1}{8}\right)s^i \tens s^i - \frac{3}{2}\left(   \frac{1}{4a^2} +\frac{\dot{a}^2}{a^2} \right) \extd t \tens \extd t,\\
       \square &= \frac{1}{a^2}\sum_i\del^2_i - 3\frac{\dot{a}}{a}\del_{t} - \del^2_t \end{align*}
   using the naive definition of the Einstein tensor. One can check that we again have $\nabla\cdot {\rm Eins}=0$.

 For a cosmological model, we take a dust stress tensor
 \[ T=p g+ (f+p)\extd t \tens\extd t= f\extd t\tens\extd t + p a(t)^2 s^i\tens s^i.\]
  This time the continuity equation $\nabla\cdot T=0$ is $\dot f+ 3(f+p){\dot a\over a}=0$
 and the Einstein equation ${\rm Eins}+4 \pi G T=0$ entails this and comes out as
 \[ f={3\over 8\pi G }\left({\dot a^2\over a^2} + {1\over 4 a^2}\right),\quad p= -{1\over 4\pi G}{\ddot a\over a}- {1\over 3} f.\]
{\em These are identical} to the corresponding equations for the classical closed 4D FLRW model with curvature constant $ 1/(4a^2_0)$ as explained in \cite{ArgMa2}. An equation of state then leads to the standard Friedmann equations for this 4D case.

\subsection{Black hole background with polygon}
Here, \cite{ArgMa2}  analyses general metrics of the static  `spherically symmetric' form with $\Z_n$ for the angular sector and shows that solving 
\[ {\rm Ricci}=0\]
leads to the following  `discrete black hole' QRG with a parameter $R_H$  of length dimension:
\begin{align*}
    g =& -{r_H\over r}\extd t \tens \extd t + {r\over r_H} \extd r \tens \extd r - r^2e^+\tens_s e^-,\\
    (\extd t,\extd t) &=-{r\over r_H} ,\quad  (\extd r,\extd r) = {r_H\over r}, \quad (e^\pm,e^\mp) = -{1 \over r^2},\\
    \nabla \extd t =&  {1\over 2r} \extd r \tens_s \extd t,\quad \nabla \extd r = -{1\over 2r} \extd r \tens \extd r - r_H e^+ \tens_s e^- + {r^2_H\over 2r^3} \extd t \tens \extd t,\\
    \nabla e^{\pm} =&   - \frac{1}{r} \extd r \tens_s e^{\pm},  \\
    {\rm R}_\nabla \extd t =& {1\over r^2} \extd t \wedge \extd r \tens \extd r + {r_H\over 2r} \extd t \wedge e^+\tens_s e^-,\\
    {\rm R}_\nabla \extd r =& -{r_H^2\over r^4} \extd r \wedge \extd t \tens \extd t
    +{r_H\over 2r}\extd r \wedge e^+ \tens_s e^-, \\
    {\rm R}_\nabla e^{\pm} =& -{1\over 2 r^2 }e^{\pm} \wedge \extd r \tens \extd r + {r_H^2\over 2r^4}e^{\pm} \wedge \extd t \tens \extd t \mp {r_H\over r} e^+ \wedge e^- \tens e^\pm,\\
   	\square =&  -{r\over r_H}\del_t^2 + {r_H\over r}\del_r^2 + {r_H\over r^2}\del_r+{2\over r^2}(\del_+ + \del_-).
\end{align*}
The radial function here reminds us of the function $1- {r_H\over r}$ for a Schwarzschild black hole with  $r<<r_H$, i.e. well inside the horizon. The comparable classical geometry with metric
\[ g = -{r_H\over r}\extd t \tens \extd t + {r\over r_H} \extd r \tens \extd r +r^2\extd\theta\tens\theta\]
 is not Ricci flat (as the Ricci tensor is sensitive to the one-higher cotangent dimension) but has vanishing Ricci scalar.

\subsection{Black hole background with fuzzy sphere}
Similarly, \cite{ArgMa2}  analyses general metrics of the static  `spherically symmetric' form with the fuzzy sphere and shows  that solving ${\rm Ricci}=0$ leads to the following  {\em fuzzy black hole} and its quantum geometry with a parameter $R_H$  of length dimension:
\begin{align*} g&= -(1-{r_H^2\over r^2})\extd t \tens \extd t + (1-{r_H^2\over r^2})^{-1} \extd r \tens \extd r + r^2 k s^i\tens s^i,\\
    (\extd t,\extd t) &=-{r^2\over r^2 - r_H^2} ,\quad  (\extd r,\extd r) = 1 - \frac{r_H^2}{r^2}, \quad (s^i,s^j) = {\delta^{ij} \over kr^2},\\
    \nabla \extd t &=- {r_H^2\over r(r^2 - r_H^2 )} \extd r \tens_s \extd t, \\
    \nabla \extd r &= {r_H^2\over r(r^2 - r_H^2)} \extd r \tens \extd r -{r_H^2\over r^3}\left(1-{r_H^2\over r^2}\right)\extd t \tens \extd t+ rk\left(1 - {r_H^2\over r^2}\right)  s^i \tens s^i, \\
    \nabla s^i &= -\frac{1}{2}\epsilon^{i}{}_{jk} s^j\tens s^k - \frac{1}{r} \extd r \tens_s s^i, \\
    {\rm R}_\nabla\extd t & = -{3r_H^2\over r^2(r^2-r_H^2)} \extd t \wedge \extd r \tens \extd r + \left( {r_H\over r} \right)^2k\extd t \wedge s^i \tens s^i,\\
    {\rm R}_\nabla\extd r & = \left( {r_H\over r} \right)^2 k\extd r \wedge s^i \tens s^i  + 3r_H^2{r^2-r_H^2\over r^6} \extd r \wedge \extd t \tens \extd t, \\
    {\rm R}_\nabla s^i & =  \left(-{1 \over 4} + k\left( 1 - {r_H^2\over r^2} \right)  \right) s^i \wedge s^j \tens s^j + \left( {r_H\over r} \right)^2{1\over r^2 - r_H^2} s^i \wedge \extd r \tens \extd r\nonumber\\&\quad + {r_H^2\over r^6}(r_H^2-r^2)s^i \wedge \extd t \tens \extd t,\\
    \square &= - \left(1 - {r_H^2\over r^2}\right)^{-1}\del_t^2 +  \left( {3\over r} - {r_H^2\over r^3} \right)\del_r + \left(  1 - {r_H^2\over r^2} \right)\del_r^2 + {1\over kr^2}\sum_i\del^2_i,\end{align*}
where
\[ k= {1\over 3}(\sqrt{7}-1) \]
is needed for ${\rm Ricci}=0$. Remarkably, the radial form here is identical to that of a classical 5D Tangherlini black hole\cite{Tang}. In both this and the preceding $\Z_n$ model, \cite{ArgMa2} begins to explore the physics using the Klein-Gordon equation for the Laplacian $\square$. By taking a quantum mechanical limit, one can see that Tangherlini radial form means that the weak field force law is no longer Newtonian gravity but has an inverse cubic form according to potential  $-{r_H^2\over 2r^2}$. This is rather different from modified gravity schemes such as MOND for the modelling of dark matter\cite{Mond}, but could still be of interest.

\subsection{Discussion of background QRG models} \label{secbackdis}

In the above models, we see the impact of the extra cotangent direction forced by the noncommutative geometry. Remarkably,  the Friedmann equations are exactly the same as for a classical geometry of one dimensional higher, and likewise for the fuzzy sphere imposing ${\rm Ricci}=0$ for a static spherically symmetric form of metric led exactly to the radial form of the 5D black hole. In short, there is a kind of `dimension jump'\cite{ArgMa2} in the radial behaviour as a result of quantising the angular coordinates.

On the other hand, these models are only mildly quantum, with noncommutativity confined to the angular sector, which allowed us to proceed much as classically. The naive the stress and Einstein tensors also worked well, where, in the absence of a theory of noncommutative variational calculus, we just went by analogy with the classical formulae. At the moment, there are no general theorems that our tensors had to be conserved, but we saw in our examples that this was the case for a natural definition of divergence. Also in the fuzzy black hole, one can set $\extd t=0$ and drop $t$ for the spatial geometry\cite{ArgMa2} and compute its nonzero Einstein tensor to find that $\nabla\cdot{\rm Eins}=0$.  Model building like this in concrete examples provides reference points to help with a general theory, which obviously should be developed. Note that $\C_q[S^2]$ has a standard 2D QRG\cite{BegMa} and hence if we take this for the angular sector then we will not have the dimension jump phenomenon. The eigenvalues of the spatial Laplacian will now be $q$ deformed, providing an even milder deformation of the classical case which could be of interest to look at.

For further work, one should really develop cosmological and black hole models with noncommutative $r$ or $t$. Such a cosmological model with only a mild quantisation is the quantum Bertotti-Robinson one in \cite{MaTao, BegMa}. No such black hole models are currently known exactly within QRG, but one was constructed \cite{Ma:alm} in an earlier `wave operator approach' (where we write down a noncommutative Laplacian directly, based on other considerations). This has the bicrossproduct model spacetime algebra (\ref{bicross}) and a  wave operator which looks like that of its flat QRG far from the horizon, but there is a new feature of a `quantum skin' just above the event horizon. We refer to \cite{Ma:alm,BegMa} for more details. Partly in this direction, one can discretise $t$ to an integer lattice $\Z$, and here particle creation was demonstrated in \cite{Ma:haw}. This should certainly be extendable to other models with discrete time. In general, quantum gravity effects\cite{MukWin} that relate to QFT on a curved background should be extendable to QRG backgrounds.

\section{Noncommutative Kaluza Klein models and dynamical mass}\label{seckk}

 This section has some new results. Motivated by Connes' idea to explain particle physics by tensoring the coordinates of a classical spacetime $M$ with a finite noncommutative geometry, we take a first look at how this could go from the QRG point of view.  Here,  $A=C^\infty(M)\tens A_f$ for some finite-dimensional algebra $A_f$, with $M$ a classical (pseudo)-Riemannian manifold e.g., Minkowski spacetime. We denote the classical curvatures of the latter with a subscript $M$.

 For the differentials, we keep $\Omega(M)$ classical and we take the tensor product exterior algebra which means $\extd x^\mu$ commute with $A_f$ and anticommute with its 1-forms. We also assume for simplicity that $\sigma(\extd x^\mu\tens (\ ))$ and $\sigma((\ )\tens \extd x^\mu)$ are the flip when the other argument is $\extd x^\nu$ or one of the basis 1-forms in $\Omega^1(A_f)$.  We assume here that the latter has a basis over $A_f$ and that these form a Grassmann algebra. In that case, the classical antisymmetric lift
\[  i(\extd x^\mu\wedge\extd x^\nu)={1\over 2} (\extd x^\mu\tens \extd x^\nu- \extd x^\nu\tens\extd x^\mu)\]
extends by the same formula when one or both of the $\extd x^\mu,\extd x^\nu$ are replaced by basic 1-forms in $\Omega^1(A_f)$.  These assumptions are all similar to those in the models of Section~\ref{secback}. 

The first model here will be carried through in detail and shows how a scalar field on $M\times\Z_n$ appears on $M$ as a multiplet with a spread of dynamically generated masses, see Corollary~\ref{corlapconst} and Proposition~\ref{propphiaction}. This establishes proof of concept, with more realistic models to be examined elsewhere.

\subsection{Type I model -- finite graph $A_f$}\label{secKKZn} Here, we again take a  Cayley graph on an Abelian group, so $A_f=C(G)$ with chosen finite group generators $a\in\CC\subseteq G\setminus\{e\}$ and basis $\{e_a\}$ which are Grassmann and have $\extd e^a=0$. But now this is tensored onto a classical spacetime with result that the general form of a metric is
\[  g= g_{\mu\nu}(x,t,i)\extd x^\mu\tens \extd x^\nu+ h_{ab}(x,t,i)e^a\tens e^b,\quad h_{ab}(x,t,i) = h_a(x,t,i) \delta_{a,b^{-1}}\]
where we indicate the functional dependence on the location $i\in G$. There are no $\extd x^\mu\tens e^a$ or $e^a\tens\extd x^\mu$ terms as the metric has to be central, and this also dictates the form of  $h_{ab}$. We assume for simplicity that this is edge-symmetric, which is 
\begin{equation} R_{a}(h_{a^{-1}})=h_a.\end{equation}

The general form of a torsion free connection is 
\begin{align}
    \nabla \extd x^\mu &= -\Gamma^\mu_{\alpha\beta} \extd x^\alpha\tens\extd x^\beta + B^\mu_{c\alpha}(\extd x^\alpha \tens e^c + e^c\tens \extd x^\alpha) + C^\mu_{ab}e^a\tens e^b \\
    \nabla e^a &= -D^a_{\alpha\beta}\extd x^\alpha\tens\extd x^\beta  + E^a_{c\alpha}(\extd x^\alpha \tens e^c + e^c\tens \extd x^\alpha) - \gamma^a_{bc}e^b\tens e^c
\end{align}
with
\begin{align}
    \label{tfc}
     D^a_{\alpha\beta}=D^a_{\beta\alpha},\quad \Gamma^\mu_{\alpha\beta}  = \Gamma^\mu_{\beta\alpha}, \quad C^\mu_{bc} = C^\mu_{cb}, \quad \gamma^a_{bc} = \gamma^a_{cb}.
\end{align}
However, to have a bimolude connection which is compatible with the commutation relations of the algebra and the differential, we also have that
\begin{align}
    \label{bim-rel}
    D^a_{\alpha\beta} = 0, \quad E^a_{c\mu} = E_{a\mu} \delta_{a,c}, \quad B^\mu_{a\nu} = 0, \quad C^\mu_{ab}= C^\mu_{a}\delta_{a,b^{-1}}.
\end{align}

Metric compatibility with this form of $\nabla$ then comes down to:
\begin{align*}
  \extd x^\alpha \tens \extd x^\beta \tens \extd x^\gamma &: \del_\alpha g_{\beta\gamma} - g_{\mu\gamma}\Gamma^\mu_{\alpha\beta} - g_{\beta\mu}\Gamma^\mu_{\alpha\gamma}= 0 \\
  \extd x^\alpha \tens \extd x^\beta \tens e^a &: 0 = 0 \\
  \extd x^\alpha  \tens e^a \tens  \extd x^\beta &: 0 = 0 \\
  e^a \tens \extd x^\alpha \tens \extd x^\beta  &: \del_a g_{\alpha\beta} = 0 \\
  e^a \tens e^b \tens \extd x^\alpha &: g_{\mu\alpha}C^\mu_{ab} + h_{mn}R_m(E^n_{l\alpha})\sigma^{ml}{}_{ab}= 0 \\
  e^a \tens \extd x^\alpha \tens e^b &: g_{\alpha\mu}C^\mu_{ab} + h_{cb}E^c_{a\alpha} = 0\\
  \extd x^\alpha \tens e^a \tens e^b   &: \del_{\alpha}h_{ab} + h_{cb}E^c_{a\alpha} + h_{ac}R_{a}(E^c_{b\alpha})= 0 \\
  e^a \tens e^b \tens e^c &: \del_{a}h_{bc} - h_{dc}\gamma^d_{ab} - h_{nd}R_{n}(\gamma^d_{mc})\sigma^{nm}{}_{ab} = 0.
\end{align*}

The 4th equation says that $g_{\mu\nu}=g_{\mu\nu}(x,t)$ is just a metric on spacetime and the 1st  then says that $\Gamma$ is its usual classical Levi-Civita connection. The 8th is the equation, at each $x,t$, for metric compatibility of $h_{ab}$ and $\gamma^a_{bc}$ and says that we have a finite QRG on $C(G)$. We have seen examples in earlier sections. The new thing we have are the fields  $E_\alpha(x,t,i)$ which we regard as a family of 1-forms on spacetime which also depend on the location $i$ in the group. The 7th equation relates this to how the internal metric $h_{ab}$ varies on spacetime. In terms of $h_a$ and given the edge-symmetry, this is
\begin{equation}
    \label{eq1}
    \del_\alpha h_a + h_a(E_{a\alpha} +R_a E_{a^{-1}\alpha}) = 0.
\end{equation}

Next, the 6th equation together with (\ref{bim-rel}) tells us
\begin{equation}
    C^\mu_{a} = - g^{\alpha\mu}E_{a\alpha} h_a
\end{equation}
Substituting this and taking into account (\ref{bim-rel}), the 5th equation becomes
\begin{equation}\label{eq2}
    h_aE_{a\alpha}\delta_{a,b^{-1}} = R_{n^{-1}}(h_n E_{n\alpha})\sigma^{n^{-1},n}{}_{a,b}
\end{equation}
with sum over $n$.

From the reality or $*$-preserving conditions we find $\Gamma^\mu_{\alpha\beta}$ has to be real and
\begin{equation}
    \label{str-pres-mZ}
    \overline{E_{a\mu}} = R_{a^{-1}}(E_{a^{-1}\mu}), \quad \overline{C^\alpha_a} = C^\mu_{a^{-1}} , \quad C^\mu_{m^{-1}}e^{m^{-1}}\tens e^m = \overline{C^\mu_{m}}\, \sigma(e^m\tens e^{m^{-1}}), 
\end{equation}
where the first two equations come from $\nabla e^a$ and the last one from $\nabla x^\mu$. As before, we use $(e^{a})^* = -e^{a^{-1}}$ and $(\extd x^{\mu})^* = \extd x^\mu$.

Our analysis so far applies to any graph. In the remainder of this section, we specialize to the case where the graph is a Cayley graph corresponding to $\Z_n$ and $A_f=C(\Z_n)$. Then the metric has the product form
\begin{equation}\label{metricZ} g=g_{\mu\nu}(x,t)\extd x^\mu\tens\extd x^\nu+\sum_\pm h_\pm (x,t,i)e^\pm\tens e^\mp,\quad h_\pm=R_\pm(h_\mp).\end{equation}
The metric on $\Z_n$ is arbitrary (edge-symmetric) as in  Proposition~\ref{QLCZn} (with $h_+$ in the role of $a$ there) but now can vary in spacetime. We denote its QRG structures at each $x,t$ with a subscript $\Z_n$. Thus,
\[ \nabla_{A_f}e^\pm:=\Gamma^\pm e^\pm\tens e^\pm;\quad \Gamma^{+} = 1-\rho,\quad  \Gamma^{-} = 1-R^2_-(\rho^{-1}),\quad \rho(x,,t,i) = {R_+(h_+)\over h_+}\] is the canonical QLC on $\Z_n$ in Proposition~\ref{QLCZn} now depending additionally on spacetime.

\begin{proposition} There is a unique $*$-preserving QLC with real coefficients for the product metric (\ref{metricZ}), namely given by
\[   \nabla \extd x^\mu = -\Gamma^\mu_{\alpha\beta} \extd x^\alpha \tens \extd x^\beta +\sum_\pm \omega^\mu  e^\pm \tens e^\mp,\quad  \nabla e^\pm = -g_{\mu\nu}\frac{\omega^\nu }{h_\pm} (e^\pm\tens_s \extd x^\mu) + \nabla_{\Z_n}e^\pm,\]
where $\Gamma^\mu_{\alpha\beta}$ is the usual classical Levi-Civita connection of $g_{\mu\nu}(x,t)$ and
\[ \omega^\mu(x,t):={1\over 2}g^{\mu\nu}\del_\nu h_\pm\]
is a vector field on spacetime and independent of the choice of $h_\pm$.
\end{proposition}
\proof The arguments are similar to those for the product metric in \cite{ArgMa1}. As consequence of  (\ref{eq1}) and (\ref{eq2}), we have $\del_{-}\del_\alpha h_a = 0$, which implies that $h_a(x,t,i) = X_a(x,t) + Y_a(i)$ for some $X_\pm,Y_\pm$. Using this and the first two equations of (\ref{str-pres-mZ}) we obtain
\[C^\mu_{\pm} = g^{\alpha\mu}{\del_\alpha h_\pm\over 2} + \imath \chi^\mu_\pm(x,t), \quad  E_{\pm\alpha} = -\frac{\del_\alpha h_{\pm}}{2h_\pm} - \imath g_{\alpha\mu}\chi^\mu_\pm(x,t).\]
For simplicity, we then ask for the coefficients of $\nabla$ to be real and hence $\chi_\pm=0$ (this is not required by the quantum geometry itself and strictly speaking the QLC is not unique, but this is a natural restriction). The condition $\del_{-}\del_\mu h_a = 0$ also tells us that $\del_\mu h_\pm$ are independent of $i$.  Even more, edge symmetry tells us that $0=\del_-\del_\mu h_+=\del_\mu(R_-(h_+)-R_+)=\del_\mu h_--\del_\mu h_+$ so that $\del_\mu h_\pm$ is independent of $\pm$. Putting all this together, we obtained the result stated. \endproof

For this connection, the Laplacian has the form
\[  \square f(x,t,i) =  g^{\alpha\beta}\del_\alpha\del_\beta f - g^{\mu\beta}\Gamma^\alpha_{\mu\beta} \del_\alpha f+ ({1\over h_{+}}+{1\over h_-})\big(-(\del_+ + \del_-)f + \omega^\mu \del_\mu f\big)  \]
which we see  is the usual Laplacian $\square_{LB}$ on spacetime and the discrete Laplacian weighted by the metric $h$, which is the QRG Laplacian $\square_{A_f}$ on $\Z_n$ for this metric and  now varies on spacetime. We also have a cross term which is the action of our vector field $\omega^\mu$ weighted by the metric.

\begin{corollary}\label{corlapconst} If $h_\pm=h(x,t)$ is a constant in $i$ (a regular polygon) and if we Fourier transform a scalar field  in the $\Z_n$ coordinate by $f(x,t,i)=\sum_{k=0}^{n-1}e^{2\pi\imath i k\over n} f_k(x,t)$ then
\[ \square f_k(x,t) =  \square_{LB} f_k+{2 e^{-\phi} }\sin^2({\pi k\over n}) f_k + g^{\mu\nu}(\del_\mu \phi)(\del_\nu f_k);\quad h=e^\phi\]
for the total space Laplacian in terms of the $\Z_n$ Fourier coefficients. \end{corollary}
\proof Here, $\square f_k$ is defined by $\sum_k e^{2\pi\imath i k\over n} \square f_k(x,t)=\square f$ if $f$ is the Fourier transform of the $f_k$ as stated. We also used $\del_\pm e^{2\pi\imath i k\over n} = (e^{\pm {2\pi\imath  k\over n}}-1)e^{2\pi\imath i k\over n}$.  \endproof

From this point of view then, a scalar field appears
as a multiplet of $n$ otherwise identical scalar fields with different masses created by a finite value of $\phi$ (and also coupled to its derivative). Indeed, a constant value of $\phi$ means we do not have the derivative interaction and just see a multiplet with modified masses.

Next,  we calculate the curvature, Ricci tensor and Ricci scalar respectively
\begin{align*}
    {\rm R}_\nabla \extd x^\mu &= {\rm R}_{\nabla_M}\extd x^\mu \\
    &\quad +\sum_{\pm}\Big(  \mp \omega^\mu  R_{\pm}(\Gamma^\mp )  e^+\wedge e^- \tens e^\mp
    \pm {g_{\nu\alpha}\over h_\pm} \omega^\mu  \omega^\alpha e^+\wedge e^-\tens \extd x^\nu \\
    &\qquad\quad + \big(\nabla_\nu \omega^\mu   - {g_{\nu\alpha}\over h_\pm} \omega^\mu  \omega^\alpha \big) \extd x^\nu \wedge e^\pm\tens e^\mp \Big)\\
    {\rm R}_\nabla e^\pm &= {\rm R}_{\nabla_{A_f}} e^\pm +\left( \nabla_\mu(g_{\nu\alpha}{\omega^\alpha\over h_\pm})  + {g_{\mu\alpha}g_{\nu\beta}\over h_\pm^2}\omega^\alpha \omega^\beta\right) e^\pm\wedge \extd x^\mu \tens \extd x^\nu \\
    &\quad    \pm {g_{\mu\nu}\over h_\pm} \omega^\mu  \omega^\nu e^+\wedge e^-\tens e^\mp+\left(\del_\mu\Gamma^\pm  +(1 - \Gamma^\pm)   g_{\mu\nu}\omega^\nu \del_\pm({1\over h_\pm})\right)\extd x^\mu\wedge e^\pm\tens e^\pm\\
    &\quad + g_{\mu\nu}\omega^\nu  \del_\mp({1\over h_\pm})( \extd x^\mu \wedge e^\mp \tens e^\pm \pm e^+\wedge e^-\tens \extd x^\mu),
\end{align*}
where we used that $\Gamma^\mu_{\alpha\beta}$ and $\omega^\mu $ are constant on $\Z_n$ and that $R_\pm(h_\mp)=h_\pm$ from edge symmetry to simplify formulae. We then recognised spacetime covariant derivatives on vector fields and 1-forms. Note that, using the metric to raise and lower indices, that $\nabla_\mu$ is the Levi-Civita connection and $\del_\mu h_\pm=2\omega_{\mu}$, one has
\begin{equation}\label{covomega} \nabla_\mu({\omega_{\nu}\over h_\pm})+ {\omega_{\mu}\omega_{\nu}\over h_\pm^2}={1\over h_\pm}\left(\nabla_\mu \omega_{\nu}- {\omega_{\mu}\omega_{\nu}\over h_\pm}\right)\end{equation}
so there is only one independent expression involving the covariant derivatives, which looks like the half-curvature tensor of a pure gauge $U(1)$ field. A potential term in $R_\nabla(e^\pm)$ which was the full curvature vanished for the same reason. Using the antisymmetric lift, we then find
\begin{align*}
    { \rm Ricci} &= {\rm Ricci}_M+ \text{Ricci}_{A_f} \\
    &\quad + {1\over 2}\sum_\pm
    \Big( (\nabla_\mu \omega^\mu   - {g_{\mu\nu}\over h_\pm}\omega^\mu  \omega^\nu)e^\pm\tens e^\mp   +(\nabla_\mu(g_{\nu\alpha}{\omega^\alpha\over h_\mp})  + {g_{\mu\alpha}g_{\nu\beta}\over h_\mp^2}\omega^\alpha \omega^\beta) \extd x^\mu \tens \extd x^\nu  \\
    &\quad\quad \quad \quad+ {g_{\mu\nu}\over h_\mp}\omega^\mu  \omega^\nu e^\mp\tens e^\mp+    g_{\mu\nu}\omega^\nu  \del_\mp({1\over h_\mp}) e^\mp\tens \extd x^\mu  -g_{\mu\nu}\omega^\nu {h_\pm\over h_\mp^2}\extd x^\mu\tens e^\pm\Big),
\end{align*}
where we recall that Ricci (and the scalar curvature) on spacetime are $-1/2$ of their usual values in our conventions. We used $R_\mp(h_\pm)=h_\mp$ from edge symmetry as well as
\[ R_\mp(\Gamma^\pm)=1-{h_\pm \over h_\mp}\]
for the connection in Proposition~\ref{QLCZn} to simplify the result. We see that Ricci has asymmetric terms. Contracting further, we have:

\begin{proposition}\label{propphiaction} The total space Ricci scalar for the $M\times \Z_n$ model is
\begin{align*}
    R &= R_M + R_{A_f}+ {1\over 2}\left({1\over h_+}+{1\over h_-}\right)\square_{LB}h_+ -{1\over 4}\left({1\over h_+^2}+{1\over h_-^2}\right) g^{\mu\nu}(\del_\mu h_+)(\del_\nu h_+)
\end{align*}
with $h_-=R_-(h_+)$.  In particular,  for a regular polygon with $h_\pm=h(x,t)$  constant on the group,  $R_{A_f}=0$ and
\[ R=R_M + \square_{LB}\phi + {1\over 2} g^{\mu\nu}( \del_\mu\phi)( \del_\nu\phi )\]
where $h=e^{\phi}$.
\end{proposition}
\proof We applied $(\ ,\ )$ which kills the last three terms of the Ricci tensor and (\ref{covomega}) to simplify. We then replace   $\omega^\mu $ in terms of $h_+$. For the special case, we then used \[{1\over h}\del_\mu h= \del_\mu\phi,\quad  {1\over h}\square_{LB}h= \square_{LB}\phi+ g^{\mu\nu}(\del_\mu\phi)(\del_\nu\phi).\]
\endproof

Integrating the $h(x,t)$ case over $M$ gives the usual Einstein-Hilbert action, zero from $\square_{LB}\phi$ as a total divergence, and the action of a massless free scalar field $\phi$. More precisely, we should integrate the scalar curvature over $\Z_n$ with respect to a measure and then, at each $x,t$, quantise the relative metric on $\Z_n$ much as in Section~\ref{secpoly}. The vacuum expectation value in this sector should eventually result in an effective metric $h(x,t)$ on spacetime which we view as $\phi$. This justifies the point of view that $\phi$ is a dynamical field for the generation of mass in Corollary~\ref{corlapconst}. It  remains to consider this more fully, as well as spinors in this model.

\subsection{Type II model -- Fuzzy sphere $A_f$} We take the fuzzy sphere but all we need from it is that $\Omega(A_f)$ has a central basis $\{s^i\}$ with Grassmann algebra and
\[ \extd s^i=-{1\over 2}\eps_{ijk}s^j\wedge s^k\]
 for some structure constants, stated here for the fuzzy sphere. We also need that $A_f$ has trivial centre. Then
\[ g= g_{\mu\nu}(x,t)\extd x^\mu\tens \extd x^\nu+  A_{i\mu}(x,t)(s^i\tens \extd x^\mu+\extd x^\mu\tens s^i) + h_{ij}(x,t)s^i\tens s^j\]
is the most general form of metric. There is no $A_f$ dependence of the coefficients as the metric has to be central. When we need the inverse metric, we will assume for convenience that $g_{\mu\nu}, h_{ij}$ are invertible with inverse matrices $g^{\mu\nu}, h^{ij}$ respectively and we also assume that the matrices
\begin{equation}\label{tildeg}  \tilde g_{\mu\nu}:=g_{\mu\nu}- A_{i\mu} h^{ij} A_{j\nu},\quad \tilde h_{ij}:=h_{ij}- A_{i\mu} g^{\mu\nu}A_{j\nu}\end{equation}
are invertible and denote their inverses by $\tilde g^{\mu\nu}$ and $\tilde h^{ij}$ respectively. Then one can show that the inverse metric bimodule inner product has values
 \[ (\extd x^\mu, \extd x^\nu) =\tilde g^{\mu\nu}, \quad (s^i,s^j) =\tilde h^{ij},\quad (\extd x^\mu, s^i) = (s^i,\extd x^\mu) = \tilde{A}^{i\mu},\]
where
\begin{align}\label{metricinvfuz}   g_{\mu\nu}\tilde A^{i\nu} + A_{j\mu}\tilde h^{ij}&=0,\quad A_{i\mu}\tilde g^{\mu\nu} + h_{ij}\tilde A^{j\nu}=0\\
 g_{\mu\gamma}\tilde{g}^{\gamma\nu} + A_{i\mu}\tilde{A}^{i\nu}&=\delta^\nu_\mu,\quad  h_{ik}\tilde h^{kj}+A_{i\mu}\tilde A^{j\mu} =\delta_i^j\end{align}
as required for the metric inverse property, with either of (\ref{metricinvfuz}) defining $\tilde A$ since $h,g$ are invertible. In terms of this, one can also write
\begin{equation}\label{invtildeg} \tilde g^{\mu\nu}= g^{\mu\nu}-g^{\mu\gamma}A_{i\gamma}\tilde{A}^{i\nu}, \quad \tilde h^{ij} =h^{ij}- h^{ik}A_{k\mu}\tilde{A}^{j\mu}.\end{equation}

The general form of a torsion-free connection under the reasonable assumption that the coefficients likewise have no $A_f$ dependence is
\begin{align}
    \nabla \extd x^\mu &= -\Gamma^{\mu}_{\alpha\beta} \extd x^\alpha\tens\extd x^\beta + B^{\mu}_{\alpha k} (\extd x^\alpha\tens s^k + s^k\tens  \extd x^\alpha)  + D^{\mu}_{kl} s^k\tens s^l \\
    \nabla s^i &= E^{i}_{\alpha\beta} \extd x^\alpha\tens\extd x^\beta + F^{i}_{\alpha k} (\extd x^\alpha\tens s^k + s^k\tens  \extd x^\alpha) + H^{i}_{kl} s^k\tens s^l
\end{align}
where
\[ \Gamma^{\mu}_{\alpha\beta} = \Gamma^{\mu}_{\beta\alpha}, \quad D^{\mu}_{kl} =  D^{\mu}_{lk}, \quad E^{i}_{\alpha\beta} = E^{i}_{\beta\alpha}, \quad H^i_{jk}-H^i_{kj}+\eps_{ijk}=0  \]
In order to be a bimodule connection, all the coefficients have to be constant in $A_f$, i.e. just functions of the time-space coordinates $x,t$.

Metric compatibility for this form of $\nabla$  comes down to:
\begin{align*}
    \extd x^\alpha\tens\extd x^\beta\tens\extd x^\gamma &: \del_\alpha g_{\beta\gamma} - g_{\mu\gamma}\Gamma^\mu_{\alpha\beta}  - g_{\beta\mu}\Gamma^\mu_{\alpha\gamma} + A_{i \gamma}E^{i}_{\alpha\beta} + A_{i\beta}E^{i}_{\alpha\gamma} = 0\\
     {\extd x^\alpha\tens\extd x^\mu\tens s^i\atop \extd x^\alpha \tens s^i \tens \extd x^\beta}\Big\} &: \nabla_\alpha A_{i\beta} + g_{\beta\mu}B^{\mu}_{\alpha i}  + A_{k\beta}F^k_{\alpha i} + h_{ki}E^{k}_{\alpha\beta} = 0\\
    s^i\tens\extd x^\alpha\tens\extd x^\beta &: g_{\mu\beta}B^\mu_{\alpha i} + g_{\alpha\mu}B^\mu_{\beta i} + A_{k\beta}F^k_{\alpha i} + A_{k\alpha}F^k_{\beta i} = 0\\
      \extd x^\alpha\tens s^i\tens s^j  &: \del_\alpha h_{ij} + A_{i\mu}B^{\mu}_{\alpha j} + A_{j\mu}B^{\mu}_{\alpha i} + h_{kj}F^k_{\alpha i} + h_{ik}F^k_{\alpha j}=  0\\
    s^i\tens\extd x^\alpha\tens s^j &: g_{\mu\alpha}D^\mu_{ij} + A_{j\mu}B^{\mu}_{\alpha i} + A_{k\alpha}H^k_{ij} + h_{jk}F^k_{\alpha i} = 0\\
    s^i\tens s^j \tens \extd x^\alpha &: g_{\mu\alpha}D^\mu_{ij} + A_{k\mu}B^{\mu}_{\alpha l}\sigma^{kl}{}_{ij} + A_{k\alpha}H^k_{ij} + h_{kl}F^k_{\alpha m}\sigma^{lm}{}_{ij} = 0\\
    s^i\tens s^j \tens s^k &: A_{m\mu}D^{\mu}_{lk}\sigma^{ml}{}_{ij} + A_{k\mu}D^{\mu}_{ij} + h_{lk}H^{l}_{ij} + h_{mn}H^{m}_{lk}\sigma^{nl}{}_{ij}  = 0.
\end{align*}
We also have reality/$*$-preserving conditions requiring $\Gamma,B,E,F$ to be real and \[  D^\mu_{kl}  = \overline{D^\mu_{mn} }\sigma^{nm}{}_{kl}, \quad H^i_{kl}  = \overline{H^i_{mn} }\sigma^{nm}{}_{kl}. \]

We see that there are two natural special cases (i) $A=0$ and (ii) $D,E=0$ where the 1st and 7th of these equations say that $\Gamma$ is the usual Levi-Civita connection  for $g$ and $H$ is the QLC for the fuzzy sphere -- which was solved in \cite{LirMa} as
\begin{equation}\label{fuzqlc} \sigma(s^i\tens s^j)=s^j\tens s^i,\quad H^i_{jk}= -{1\over 2} h^{im}(2\eps_{mkl}h_{lj}+ {\rm Tr}(h)\eps_{mjk}) \end{equation}
on noting that our $H^i{}_{jk}$ is $-{1\over 2}\Gamma^i{}_{jk}$ in \cite{LirMa}. In either case, things simplify, while more generally we see that the presence of $A,D,E$ means that spacetime and $A_f$ no longer have their separately metric compatible connections.

We,  next compute the Laplacian for the general case as
\begin{align*}
    \square f &=\tilde g^{\mu\nu}\nabla_\mu\del_\nu f+ (\tilde{h}^{ij}\del_j\del_i f + \tilde{h}^{kl}H^i_{kl}\del_if) \\
    &\quad +
    (\tilde{g}^{\alpha\beta}E^i_{\alpha\beta} + 2\tilde{A}^{\alpha k}F^i_{\alpha k})\del_i f +  2\tilde{A}^{\mu i}\del_\mu\del_i f+ ( 2\tilde{A}^{\alpha k}B^\mu_{\alpha k} + \tilde{h}^{kl}D^\mu_{kl})\del_\mu f,
  \end{align*}
where we recognise the first term as a Laplacian on $M$ modified to use $\tilde g^{\mu\nu}$ and the second expression as a Laplacian on the fuzzy sphere modified to use $\tilde h^{ij}$. These become respectively $\square_{LB}$ and the fuzzy sphere Laplacian for $h_{ij}$ in the diagonal case $A=0$. We see that there are also further couplings to the derivatives of $f$ in both spacetime and fuzzy directions. If $h_{ij}=h(x,t)\delta_{ij}$ is a multiple of the round metric then at each $x,t$, $\square_{A_f}$ is a multiple of the action of the quadratic Casimir of $U(su_2)$ and latter has eigenvalues labelled by $l\in \N$ in a decomposition into noncommutative spherical harmonics\cite{ArgMa2}. Hence we have a similar phenomenon as in Corollary~\ref{corlapconst}. If we use the reduced fuzzy sphere at $\lambda_P=1/n$ then we have again finite multiplets with varying mass within a multiplet.

Next, the curvature of the general torsion free connection is
\begin{align*}
    \text{R}_\nabla s^i &= (\nabla_\gamma E^i_{\alpha\beta}  - F^i_{\gamma k}E^k_{\alpha\beta})\extd x^\gamma \wedge \extd x^\alpha \tens \extd x^\beta \\
    &\quad + (\nabla_\gamma F^i_{\alpha k} - E^i_{\gamma\beta}B^\beta_{\alpha k} - F^i_{\gamma l}F^l_{\alpha k}) \extd x^\gamma \wedge \extd x^\alpha \tens s^k \\
    &\quad+ (-\nabla_\gamma F^i_{\alpha k} + E^i_{\gamma\beta}B^\beta_{\alpha k} + F^i_{\gamma l}F^l_{\alpha k}  - H^i_{kl}E^l_{\gamma\alpha}) s^k \wedge \extd x^\gamma \tens \extd x^\alpha\\
    &\quad+ (\del_\gamma H^i_{kl} - E^i_{\gamma\beta}B^\beta_{kl} - F^i_{\gamma m}H^m_{kl} + F^i_{\alpha k}B^\alpha_{\gamma l} + H^i_{kl}F^l_{\gamma l}) \extd x^\gamma \wedge s^k \tens s^l\\
    &\quad- (F^i_{\alpha k}B^\alpha_{\beta l} + H^i_{km}F^m_{\beta l} + {1\over 2}F^{i}_{\beta j}\eps^{j}{}_{kl}) s^k \wedge s^l \tens \extd x^\beta\\
    &\quad- (F^i_{\alpha k}D^\alpha_{ml} + H^i_{kn}H^n_{ml} + {1\over 2 }H^i_{nl}\eps^n{}_{km}) s^k \wedge s^m \tens s^l, \\
    \text{R}_\nabla \extd x^\mu &= (-\del_\gamma \Gamma^\mu_{\alpha\beta}-\Gamma^\mu_{\gamma\nu}\Gamma^\nu_{\alpha\beta} - B^\mu_{\gamma k}E^k_{\alpha\beta})\extd x^\gamma \wedge \extd x^\alpha \tens \extd x^\beta \\
    &\quad+ (\nabla_\gamma B^\mu_{\alpha k}  - B^\mu_{\gamma l}F^l_{\alpha k}) \extd x^\alpha \wedge \extd x^\gamma \tens s^k \\
    &\quad+ (\nabla_\alpha B^\mu_{\beta k} + B^\mu_{\alpha l}F^l_{\beta k}  - D^\mu_{kl} E^l_{\alpha\beta}) s^k \wedge \extd x^\alpha \tens \extd x^\beta\\
    &\quad+ (\nabla_\alpha D^\mu_{kl}  - B^\mu_{\alpha m}H^m_{kl} + B^\mu_{\gamma k} B^\gamma_{\alpha l} + B^\mu_{km} F^m_{\alpha l}) \extd x^\alpha \wedge s^k \tens s^l\\
    &\quad- (B^\mu_{\alpha k }B^\alpha_{\gamma l} + D^\mu_{km}F^m_{\gamma l} + {1\over 2}B^\mu_{\gamma m}\eps^{m}{}_{kl}) s^k \wedge s^l \tens \extd x^\gamma\\
    &\quad- (B^\mu_{\alpha k}D^\alpha_{ml} + D^\mu_{kn} H^n_{ml} + {1\over 2}D^\mu_{nl}\eps^{n}_{km}) s^k \wedge s^m \tens s^l.
\end{align*}
One can see within here the curvature of $\Gamma$ as a classical connection and the curvature of $H$ as a connection on the fuzzy sphere with constant coefficients on the fuzzy sphere.

We next compute the Ricci tensor (the details of which we omit) resulting in  the Ricci scalar curvature
\begin{align*}
    R=&\tilde R_M+ \tilde R_{A_f} - {\tilde g^{\alpha\beta}\over 2}(  \nabla_\alpha F^i_{\beta i} - F^i_{\alpha l}F^l_{\beta i}  + E^k_{\alpha\beta} B^\mu_{\mu k} - E^i_{\alpha\mu}B^\mu_{\beta i} - E^k_{\mu\beta} B^\mu_{\alpha k} + H^i_{il}E^l_{\alpha\beta} )\\
    & \quad \quad\quad\quad\quad - {\tilde h^{kl}\over 2}( B^\mu_{\mu m}H^m_{kl}- B^\mu_{\gamma k} B^\gamma_{\mu l} -
    \nabla_\mu D^\mu_{kl} + F^i_{\alpha i}D^\alpha_{kl}  -F^i_{\alpha k}D^\alpha_{il}     - B^\mu_{km} F^m_{\mu l})
\end{align*}
where
\begin{align*} \tilde R_M&={\tilde g^{\alpha\beta}\over 2}(\del_\alpha \Gamma^\mu_{\mu\beta}-\del_\mu \Gamma^\mu_{\alpha\beta}+\Gamma^\mu_{\alpha\nu}\Gamma^\nu_{\mu\beta}-\Gamma^\mu_{\mu\nu}\Gamma^\nu_{\alpha\beta})\\
 \tilde R_{A_f}&={\tilde h^{kl}\over 2} ( H^i_{kn}H^n_{il}-H^i_{in}H^n_{kl}  - H^j_{pl}\eps^p{}_{jk} )
 \end{align*}
reduce to the usual Ricci tensor on spacetime (in our conventions) and to the $R_{A_f}$  on the fuzzy sphere\cite{LirMa} as used in Section~\ref{secfuzzy} in the diagonal case where $A=0$.

We now look at the case where $D=E=0$ but $A$ is not necessarily zero. In this case, the content of the above metric-compatibility equations is as follows. The 1st and 7th metric compatibility equations reduce again to $\Gamma$ the Levi-Civita connection for $g$ and $H$ the QLC for  $h$.  The remaining fields $B,F$ in the quantum Levi-Civita connection are uniquely determined from the 2nd and 5th/6th  metric compatibility equations, which can be solved as
\begin{equation}\label{Bsoln} \tilde g_{\mu\alpha}B^\alpha_{\nu i}=-\nabla_\nu A_{i\mu}+A_{j\mu}A_{k\nu}H^j_{im}h^{km},\end{equation}
\begin{equation}\label{Fsoln} F^i_{\mu j}=-A_{k\mu}H^k_{jm}h^{im}- h^{ik}A_{k\nu}B^\nu_{\mu j}\end{equation}
and the remaining 3rd, 4th equations reduce respectively to
\begin{equation}\label{DEqlc}\nabla_\mu A_{i\nu}+\nabla_\nu A_{i\mu}=0,\quad \CD_\mu\tilde h_{ij}=0\end{equation}
as conditions on the extended metric for the QLC to exist.  We define
\[  \CD_\mu  f_{ij}:=\del_\mu  f_{ij}+ f_{kj}F^k_{\mu i}+ f_{ik}F^k_{\mu j}\]
as the covariant derivatives  on a matrix-valued field scalar $f_{ij}$ with respect to $F^i_{\mu j}$ as a matrix valued gauge field for the internal (roman) indices. If $F$ is known then $B^\mu_{\nu i}$ can also be written similarly in terms of an extended $\CD_\nu A_{i\mu}$. Solutions of these remaining equations will be considered elsewhere, but one is of course $A=0$ and $h_{ij}$ constant in spacetime.

This describes the extended quantum geometry, but we still need to connect it to the Kaluza-Klein point of view. The key observations are:
\begin{itemize}
\item We should regard $\tilde g_{\mu\nu}$ as the physically observed metric for GR and 
\[  g_{\mu\nu}=\tilde g_{\mu\nu}+ A_{i\mu} h^{ij} A_{j\nu} \]
from  (\ref{tildeg}) as the `Kaluza-Klein ansatz'. 
\item When the above Ricci scalar is expressed in terms of the Ricci scalar of the metric $\tilde g$, what is left should be $R_{A_f}$ and something resembling the Maxwell or Yang-Mills action of $A_{i\mu}$ as some kind of connection. This is a long computation, which will not be attempted here.
\item Note, however,  that the lowered $B$ in (\ref{Bsoln}) when antisymmetrized looks a lot like a Yang-Mills curvature with `Lie algebra' structure constants $H^j_{im}h^{km}-H^k_{im}h^{jm}$  built from the QRG of $A_f$. 
\item The Kaluza-Klein `cylinder assumption' that the coefficients of the extended geometry are constant in the extra directions for us is a consequence of centrality of the metric  and the trivial centre of $A_f$, i.e. comes out of the quantum geometry.
\item The case of the round metric $h_{ij}=h(x,t)\delta_{ij}$ on the fuzzy sphere has a single dilation field $h(x,t)$ as in usual Kaluza-Klein theory. Here,  $H^i_{jk}=-{1\over 2}\eps_{ijk}$ so that the `Lie algebra' structure constants suggested by (\ref{Bsoln}) are those of $su_2$. 
\end{itemize}
We see that the Kaluza-Klein idea with the fuzzy sphere replaces the scalar field by a matrix valued function $h_{ij}$, the `gauge field' $A_{i\mu}$ has an internal index, there is an induced matrix-valued gauge field $F^i{}_{\mu j}$ built from $A_{i\mu}$, and there are restrictions on this data coming out of the quantum geometry which we analysed for the simpler $D=E=0$ case as (\ref{DEqlc}). The full development of this model remains for further work. 

\section{Concluding remarks}\label{seccon}

With these notes, we hope to have convinced the reader that noncommutative geometry can be used for quantum gravity both for baby quantum gravity model-building and to model putative quantum gravity corrections to spacetime geometry under the quantum spacetime hypothesis. While noncommutative geometry itself has a long history and several approaches, the most well-known being the one of Connes\cite{Con}, we have adopted a more constructive approach that grew out of models with quantum group symmetry in the 1990s in works such as \cite{Ma:rie} and for which there is now a systematic treatment \cite{BegMa}. This text covers the mathematics of QRG but stops short of quantum gravity itself. The reason is that the issues for that are to do with the very nature of the Ricci tensor, the variational principle etc., all of which need a deeper and more abstract understanding of the physics before we can confidently  transfer them over to the quantum case. A recent step towards noncommutative variational calculus here is a  theory of quantum jet bundles\cite{MaSim}.  The QRG formalism may also need to be extended e.g. to include metrics with a weaker notion of inverse (so as to not be forced to be central) or to step back to more general connections, for example using an earlier frame bundle approach\cite{Ma:rie}. But none of this should stop us meanwhile, as we have seen, from already feeling our way in `model building' and starting to study physics on such models using the current formalism as base.

In terms of such model building, while we have come a long way since early flat quantum spacetime models, we still have to address the question of how exactly do we go from noncommutative algebra to physical interpretations. Ideas such as normal ordering for the identification with classical waves\cite{AmeMa} were model-specific in the absence of a systematic approach. In QGQG, an answer is provided by the functional integral approach. Here the QRG is confined to the action with a map $\int: A\to \C$ converting the scalar curvature to an actual number. We can similarly compute correlation functions in other  functional integral QFTs, for example on $\Z$  in \cite{Ma:haw}. But if we want to think about physics directly on the QRG, we need new tools to face the fundamental problem of how to think physically about a noncommutative coordinate algebra $A$. For example, if we take a quantum mechanical view then what is the role of the time with respect to which the quantum mechanics is defined?

Recent progress on this front is the notion of quantum geodesics\cite{Beg:geo,BegMa:mec,BegMa:cur,LiuMa}. This is mathematically challenging (it uses a theory of $A$-$B$-bimodule connections where $B$ is the coordinate algebra for the geodesic time), and it also challenges our physical intuition when applied to quantum spacetime. If $A$ refers only to space then the idea is quite simple: instead of evolving one geodesic at a time we consider a kind of fluid of particles evolving in time, or more precisely we evolve a wave function $\psi$ where classically $|\psi|^2$ is the probability density in a quantum mechanical picture. All the particle tangent vectors fit together to a global velocity field which also evolves, and all of this extends nicely to a general QRG\cite{Beg:geo,BegMa:cur}. In the general case, $\psi^*\psi$ is a positive element of $A$ but one can apply a positive linear functional $\int:A\to \C$ on $A$ as a quantum system to render evolving expectation values. Thus, the physical probabilities are a composite of an evolving geodesic `wave function' $\psi\in A$ and a fixed state  on the algebra in a $*$-algebra sense. When $A$ refers to spacetime, however, the classical picture has $\psi$ a wave-function on spacetime (so $|\psi|^2$ is a probability density for the particle location in spacetime, which is somewhat unfamiliar) and the geodesic flow time parameter is external to spacetime and therefore  represents a role for an observer. Here, \cite{LiuMa} explores these ideas for classical Minkowski space and then for the bicrossproduct or $\kappa$-Minkowski spacetime (\ref{bicross}) with its flat QRG, with some principal findings as follows. (i) On a classical spacetime, we can take $\psi$ real and then the theory is equivalent to ordinary geodesics done in a fluid-like way. But on the quantum spacetime, initially real $\psi$ evolve with complex corrections. These can then have interference effects as in quantum mechanics.
(ii) If one tries to model a point in spacetime as a `bump function' then one finds quantum corrections which blow up as the width approaches $\lambda_P$, in keeping with the idea that the continuum is not a valid concept at the Planck scale.

Beyond the immediate future, one would of course like any theory of quantum gravity to answer current puzzles such as the size of the cosmological constant, the link between entropy and geometry evident from black holes (see, e.g. \cite{Suss} and recent ideas in \cite{Wit,Don}), state reduction\cite{Pen}, the structure of particle masses etc. While we are still some way from this, we saw in Section~\ref{secKKZn}  how tensor product by a finite geometry  can generate multiplets of fields with different masses, as proof of concept towards solving the `generations problem' and, ultimately, an explanation of particle masses. Our approach differs from \cite{ConMar} but should be extended to include spinors and to QRGs offering a better fit.

We also note that ideas of quantum computing such as the Kitaev model\cite{Kit} already have a lot in common with TQFT and hence with 2+1 quantum gravity (see \cite{CowMa} for some recent work on the structure of this model), albeit not yet linked to QRG. In 2+1 quantum gravity the q-deformed version corresponds to switching on a cosmological constant while \cite{ArgMa3} shows that q-deformation of the QRG arises naturally from truncation of $\N$ to a finite graph $\bullet$-$\bullet$-$\cdots$-$\bullet$, possibly hinting at a different point of view on its physical necessity. The cosmological constant was also posited in \cite{MaTao} to be a consequence of quantum spacetime (and small for this reason) but without a proposal for a mechanism. Moreover, the truncation from $\Z$ to $\N$ in \cite{ArgMa3} forces the QRG to have a direction dependence that decays far from the first node $1\in \N$, suggesting radically new physics emanating from this boundary into the bulk.

More generally, the language of quantum information provides new tools for quantum gravity. Some related ideas for the role of QRG in quantum computing are in \cite{Ma:log}. It is also possible, in principle, to build quantum Riemannian geometries and quantum groups into silicon chips, by working over $\F_2=\{0,1\}$ in place of $\C$, see \cite{MaPac,MaPac1}. At the moment,  this provides a unique glimpse of the total moduli space of `everything out there' in low dimension, which is insightful even if we are ultimately interested in working over $\C$. This can also be a useful technique if calculations over $\C$ become intractable, to first look at them over finite fields. Over $\F_2$, one also gets a Venn diagram picture of the exterior differential and can explore novel ideas related to de Morgan duality\cite{Ma:boo}.


\begin{thebibliography}{99}
    \bibitem{Loll}J. Ambjorn, J. Jurkiewicz and R.  Loll, Dynamically triangulating Lorentzian quantum gravity. Nucl. Phys. B, 610 (2001) 347--382
    \bibitem{AmeMa}G. Amelino-Camelia and S. Majid, Waves on noncommutative spacetime and gamma-ray bursts, Int. J. Mod. Phys. A 15 (2000) 4301--4323
    \bibitem{ArgMa1} J.N. Argota-Quiroz and S. Majid, Quantum gravity on polygons and $\R\times\Z_n$ FLRW model,  Class. Quantum Grav. (2020) 245001 (43pp)
    \bibitem{ArgMa2} J.N. Argota-Quiroz and S. Majid, Fuzzy and discrete black hole models, Class. Quantum Grav. 38 (2021) 145020 (36pp)
     \bibitem{ArgMa3} J.N. Argota-Quiroz and S. Majid, Quantum Riemannian geometry of the discrete interval and q-deformation, arXiv:2204.12212 (math.QA)
    \bibitem{Ash}A. Ashtekar, T. Pawlowski, P. Singh and K. Vandersloot, Loop quantum cosmology of k = 1 FRW models. Physical Review D,  75 (2007) 024035
    \bibitem{BatMa}E. Batista and S. Majid, Noncommutative geometry of angular momentum space $U(su_2)$, J. Math. Phys. 44 (2003) 107--137
    \bibitem{Beg:geo}E.J. Beggs, Noncommutative geodesics and the KSGNS construction, J. Geom. Phys. 158 (2020) 103851
    \bibitem{BegMa:gra} E.J. Beggs and S. Majid, Gravity induced by quantum spacetime, Class. Quantum Grav. 31 (2014) 035020 (39pp)
    \bibitem{BegMa:poi}E.J. Beggs and S. Majid, Poisson-Riemannian geometry, J. Geom. Phys.  114 (2017) 450--491
    \bibitem{BegMa} E.J. Beggs and S. Majid, {\em Quantum Riemannian Geometry}, Grundlehren der mathematischen Wissenschaften, Vol. 355, Springer (2020) 809pp.
     \bibitem{BegMa:mec}E.J. Beggs and S. Majid, Quantum geodesics in quantum mechanics, arXiv:1912.13376 (math-ph)
     \bibitem{BegMa:cur}E. Beggs and S. Majid, Quantum geodesics and curvature, arXiv: 2201.08244 (math.QA)
     \bibitem{Birrel} N.D. Birrell and P.C.W. Davies, {\em Quantum Fields in Curved Space}, Cambridge University Press (1984)
    \bibitem{Sit}A. Bochniak, A. Sitarz and P. Zalecki, Riemannian geometry of a discretized circle and torus, SIGMA 16 (2020), 143 (28pp)
    \bibitem{Bre}K. Bresser, F. M\"uller-Hoissen, A. Dimakis and A. Sitarz, Noncommutative geometry of finite groups. J. Phys. A, 29 (1996) 2705--2735
    \bibitem{Con}A. Connes, {\em Noncommutative Geometry}, Academic Press (1994)
    \bibitem{ConMar} A. Connes and M. Marcolli, {\em Noncommutative Geometry, Quantum Fields and Motives} (AMS Colloquium Publications Vol 55), Hindustan Book Agency, 2008.
    \bibitem{CowMa}A. Cowtan and S. Majid, Quantum double aspects of surface code models, J. Math. Phys. 63 (2022)
    \bibitem{DFR}S. Doplicher, K. Fredenhagen and J. E. Roberts, The quantum structure of spacetime at the Planck scale and quantum fields, Commun. Math. Phys. 172 (1995) 187--220
    \bibitem{Don}W. Donnelly, Y. Jiang, M. Kimb, G. Wong, Entanglement entropy and edge modes in topological string theory. Part I. Generalized entropy for closed strings, JHEP (2021) 201
    \bibitem{Dow}F. Dowker, Introduction to causal sets and their phenomenology. General Rel. and Grav. 45 (2013) 1651--1667
    \bibitem{DVM}M. Dubois-Violette and P.W. Michor, Connections on central bimodules in noncommutative differential geometry, J. Geom. Phys. 20 (1996) 218--232
       \bibitem{FreMa}L. Freidel and S. Majid, Noncommutative harmonic analysis, sampling theory and the Duflo map in 2+1 quantum gravity, Class. Quant. Gravity 25 (2008) 045006 (37pp)
    \bibitem{Hal}M. Hale, Path integral quantisation of finite noncommutative geometries. J. Geom.  Phys.  44 (2002) 115--128
    \bibitem{Hoo}G. 't Hooft, Quantization of point particles in 2 + 1 dimensional gravity and space- time discreteness Class. Quantum Grav. 13 (1996) 1023
    \bibitem{Kit}A. Kitaev, Fault-tolerant quantum computation by anyons, Ann. Phys. 303 (2003) 3--20
\bibitem{LirMa} E. Lira-Torres and S. Majid, Quantum gravity and Riemannian geometry on the fuzzy sphere,  Lett. Math. Phys. (2021) 111:29 (21pp)
    \bibitem{LiuMa}C. Liu and S. Majid, Quantum geodesics on quantum Minkowski spacetime, in press J. Phys. A (2022)
       \bibitem{Luk}J. Lukierski, H. Ruegg, A. Nowicki and V.N. Tolstoi, Q-deformation of Poincare algebra, Phys. Lett. B 264 (1991) 331
    \bibitem{Ma:pla}S. Majid, Hopf algebras for physics at the Planck scale, Class. Quantum Grav. 5 (1988) 1587--1607
    \bibitem{Ma:qreg}S. Majid, On q-regularization, Int. J. Modern Physics A. 5 (1990) 4689--4696
    \bibitem{Ma:dua}S. Majid, Duality principle and braided geometry, in Springer Lect. Notes Phys. 447 (1995) 125--144
    \bibitem{Ma:rie} S. Majid, Quantum and braided group Riemannian geometry, J. Geom. Phys. 30 (1999) 113-146
    \bibitem{Ma:non} S. Majid, Riemannian geometry of quantum groups and finite groups with nonuniversal differentials, Commun. Math. Phys. 225 (2002) 131--170
    \bibitem{Ma:alm} S. Majid, Almost commutative Riemannian geometry: wave operators, Commun. Math. Phys. 310 (2012) 569--609
    \bibitem{Ma:gra}S. Majid, Noncommutative Riemannian geometry of graphs, J. Geom. Phys. 69 (2013) 74--93
    \bibitem{Ma:haw}S. Majid, Quantum Riemannian geometry and particle creation on the integer line, Class. Quantum Grav. 36 (2019) 135011 (22pp)
    \bibitem{Ma:sq}S. Majid, Quantum gravity on a square graph,   Class. Quantum Grav 36 (2019) 245009 (23pp)
 \bibitem{Ma:log}S. Majid, Quantum geometry, logic and probability, Phil. Prob. Sci. (Zag. Fil. Nauce) 69 (2020) 191-236
    \bibitem{Ma:boo}S. Majid, Quantum geometry of Boolean algebras and de Morgan duality, in press J. Noncomm. Geom. (2022) 37pp
      \bibitem{MaOse} S. Majid and P.K. Osei, Quasitriangular structure and twisting of the 2+1 bicrossproduct model, J. High Energ. Phys. 1 (2018) 147 (28pp)
    \bibitem{MaPac}S. Majid and A. Pachol, Digital finite quantum Riemannian geometries, J. Phys. A 53 (2020) 115202 (40pp)
    \bibitem{MaPac1}S. Majid and A. Pachol, Digital quantum groups, J. Math. Phys. 61 (2020) 103510 (34pp)
    \bibitem{MaRue} S. Majid and H. Ruegg, Bicrossproduct structure of the $\kappa$-Poincar\'e group and non-commutative geometry, Phys. Lett. B. 334 (1994) 348--354
    \bibitem{MaSch} S. Majid and B. Schroers, q-Deformation and semidualisation in 3D quantum gravity, J. Phys A 42 (2009) 425402 (40pp)
    \bibitem{MaSim} S. Majid and F. Simao, Quantum jet bundles, arXiv: 2202.03067 (math.QA)
    \bibitem{MaTao}S. Majid and W.-Q. Tao, Cosmological constant from quantum spacetime, Phys. Rev. D  91 (2015)  124028 (12pp)
 \bibitem{Mond}M. Milgrom, A modification of the Newtonian dynamics as a possible alternative to the hidden mass hypothesis, Astrophysical J. 270 (1983) 365--370
    \bibitem{MukWin}V. Mukhanov and S. Winitzki, {\em Introduction to Quantum Effects in Gravity}, Cambridge University Press (2007)
    \bibitem{Pen}R. Penrose, On gravity's role in quantum state reduction, General Relativity and Gravitation 28 (1996) 581--600
    \bibitem{Sny}H.S. Snyder, Quantized space-time, Phys. Rev. 71 (1947) 38--41
    \bibitem{Suss}L. Susskind and J. Uglum, Black hole entropy In canonical quantum gravity and superstring theory, Phys. Rev. D50 (1994) 2700--11
    \bibitem{Tang}F. R. Tangherlini, Schwarzschild field in n dimensions and the dimensionality of space problem. Nuovo Cimento 27 (1963) 636--651
    \bibitem{Wit}E. Witten, Gravity and the crossed product, arXiv:2112.12828 (hep-th)
\bibitem{Wor} S.L. Woronowicz, Differential calculus on compact matrix pseudogroups (quantum groups).
Commun. Math. Phys. 122 (1989) 125--170
\end{thebibliography}
\end{document}